# Deep Cerebellar Nuclei Segmentation via Semi-Supervised Deep Context-Aware Learning from 7T Diffusion MRI

Jinyoung Kim[a], Remi Patriat[a], Jordan Kaplan[a], Oren Solomon[a], and Noam Harel[a,b]

a. Center for Magnetic Resonance Research, University of Minnesota, Minneapolis, MN, USA
b. Department of Neurosurgery, University of Minnesota, Minneapolis, MN, USA

**Corresponding Author**

Dr. Jinyoung Kim

E-mail: kimx2469@umn.edu

Phone: +1-612-309-3289

# ABSTRACT

Deep cerebellar nuclei are a key structure of the cerebellum that are involved in processing motor and sensory information. It is thus a crucial step to accurately segment deep cerebellar nuclei for the understanding of the cerebellum system and its utility in deep brain stimulation treatment. However, it is challenging to clearly visualize such small nuclei under standard clinical magnetic resonance imaging (MRI) protocols and therefore precise segmentation is not feasible. Recent advances in 7 Tesla (T) MRI technology and great potential of deep neural networks facilitate automatic patient-specific segmentation. In this paper, we propose a novel deep learning framework (referred to as DCN-Net) for fast, accurate, and robust patient-specific segmentation of deep cerebellar dentate and interposed nuclei on 7T diffusion MRI. DCN-Net effectively encodes contextual information on the patch images without consecutive pooling operations and adding complexity via proposed dilated dense blocks. During the end-to-end training, label probabilities of dentate and interposed nuclei are independently learned with a hybrid loss, handling highly imbalanced data. Finally, we utilize self-training strategies to cope with the problem of limited labeled data. To this end, auxiliary dentate and interposed nuclei labels are created on unlabeled data by using DCN-Net trained on manual labels. We validate the proposed framework using 7T B0 MRIs from 60 subjects. Experimental results demonstrate that DCN-Net provides better segmentation than atlas-based deep cerebellar nuclei segmentation tools and other state-of-the-art deep neural networks in terms of accuracy and consistency. We further prove the effectiveness of the proposed components within DCN-Net in dentate and interposed nuclei segmentation.

# I. INTRODUCTION

The cerebellum is crucially not only involved in complex motor, cognitive and linguistic tasks [1] but also emotional and perceptual processing [2], [3]. Of the cerebellum system, deep cerebellar nuclei (DCN) integrate signals from cerebellar cortex and relay them to cerebral cortex or brainstem. Therefore, the DCN play a central role to form a feedback loop of cerebellar cortex and cerebral cortex [3]. The DCN are divided into three parts: fastigial, interposed (globose and emboliform sub-nuclei), and dentate nuclei [4].

There have been on-going efforts to investigate the functional role of the DCN in neurological disorders and treatment [4]–[7]. Also, it has been reported that deep brain stimulation (DBS) therapy of the dentate nucleus is effective for post-stroke motor impairments, tremor, and cerebellar ataxia [6]–[8]. Clear visualization of the DCN is thus a pre-requisite for such neuroimaging studies or neuro-modulation planning [4]. Moreover, automatic segmentation facilitates subsequent analysis in terms of consistency and efficiency. Although there have been studies to visualize the DCN with different magnetic resonance imaging (MRI) protocols [4], [9] or provide cerebellum atlases on common templates [10], [11], it is not trivial to precisely localize such small nuclei on subject-specific standard clinical MRIs. Indeed, little work has been done to automatically segment the DCN [11], [12]. Diedrichsen et al. [11] generates a probabilistic atlas (including the DCN) on a cerebellum template (SUIT) [10] and normalizes the cerebellum anatomy of a specific subject onto the SUIT atlas space. An estimated warp deformation field is then inversely applied to the SUIT atlas for segmentation of the cerebellum in the subject space. Ye et al. [12] utilizes a geometric deformable model with a tractography initialization. Only dentate segmentation results, however, are provided on a very small set of 3 Tesla (T) diffusion MRIs. More recently, Carass et al. [13] compares state-of-the-art segmentation methods that participated in a Cerebellum Parcellation Challenge of MICCAI 2017. However, lobes, vermis, lobules, and their subcomponents in a hierarchical level of the cerebellum were the primary interest of these methods.

With the advent in ultrahigh-field MR technologies, 7T MR imaging allows clear visualization of anatomical structures due to its superior contrast and resolution [14]–[16]. More recently, the 7T MRI system (The Magnetom Terra, Siemens Medical Solutions) received a 510k cleared for clinical use by the Food and Drug Administration (FDA). A number of studies leverage 7T MRIs for functional imaging study of the DCN [17], [18]. Diedrichsen et al. [17] provides a high quality atlas of the DCN and improves a normalization process using multiple contrast 7T MRIs. Thürling et al. [18] uses 7T functional MRIs to study the activation of the dentate nucleus in a verb generation task. Visual benefits of the 7T MRI also facilitate segmentation of the DCN that appear as hypo-intense or hyper-intense. Manual delineation of the DCN is, however, time-consuming and requires the anatomical expert-knowledge that is subjective and thus prone to intra- and inter-rater variability. 7T MR atlas-based segmentation automates the procedures, but does not adequately take into account inter-subject variability and oftentimes entails additional refinement steps. Therefore, it is required to fast, accurately, and consistently segment the DCN on the subject-specific image in a fully automatic way.

During the last decade, convolutional neural networks (CNN) have re-gained much attention due to state-of-the-art performance in computer vision and image processing tasks with increasingly available training data and computational power [19]. However, since it was typically designed for classification tasks on 2D images, volumetric segmentation in medical images has been limited due to low efficiency.

Recently, fully convolutional networks (FCN) have handled the issue by considering a whole network as a large convolution filter, trained in an end-to-end manner [20]. Given images with arbitrary size, this enables dense inference in a single step, and thus redundant convolutions and pooling operations can be avoided. Therefore, the FCN and its variants have proven their effectiveness in various medical image segmentation tasks [21]–[23]. Among such architectures, U-Net [23] is the most notable approach. The key feature, a skip connection allows the network to prevent loss of contextual information at multiple image scales and thus it has shown great potential for semantic segmentation. Schlemper et al. [24] extends U-Net by exploiting an attention mechanism to learn where to focus on for medical image segmentation. However, such FCNs usually have millions of parameters and require a large amount of labeled data to avoid over-fitting [25], [26]. More recently, densely connected convolutional neural networks [27] have been developed to address those challenges and also incorporated into the FCN framework (FC-DenseNet) [25]. Dense connection enables an efficient gradient propagation, deep supervision, and reuse of features [26], [27]. The number of parameters to be optimized in the network is therefore dramatically reduced without harming its performance, which is applicable in clinical scenarios with limited labeled data.

In this paper, we propose a novel deep learning framework for fast, accurate, and robust patient-specific segmentation of the DCN (named DCN-Net). We use B0 images from 7T diffusion weighted imaging (DWI) to segment the DCN since the small nuclei appear as hypo-intense regions in these images, furthermore, DWI is becoming increasingly part of the standard protocol in the clinical workflow [28], [29]. We solely focus on dentate and interposed nuclei due to the lack of visible contrast of the fastigial nucleus in the B0 images. Fig. 1 visualizes the 7T DWI B0 image around the dentate and interposed nuclei. Since automatic segmentation mostly relies on the image appearance, it is not an easy task to simultaneously segment such small and adjacent structures, especially with low contrast boundaries, isointense, and fuzzy borders as displayed in Fig. 1. This is of significant importance in clinical scenarios where high quality data is not always available.

We herein address common yet important issues in deep learning-based segmentation of small, adjacent, imbalanced, and isointense structures - capturing contextual information on small patch images, handling highly imbalanced class labels, and overcoming the problem of limited labeled data.

3D patches (sub-volumes) based processing within the FCN is typically considered to reduce the memory burden and to significantly increase the number of training samples, especially in volumetric medical image segmentation [30]. However, learning features in very deep networks using such patches may not preserve local details of small dentate and interposed nuclei (whose volumes are approximately 630mm$^3$ and 50mm$^3$, respectively) due to consecutive pooling operations. Several networks address this issue by employing a dilated convolution which adjusts the field of view of convolutional filters, thereby capturing the contextual information without max-pooling [31]–[35]. Chen et al. [32] uses the dilated convolution with different rates in parallel to encode multi-scale contextual information. It is later extended in [35] by adding a decoder to recover boundary details. Also, Gu et al. [33] introduces a dense atrous convolution block to extract high level features.

Data imbalance problem (i.e., a large volume difference between dentate and interposed nuclei in this work) is a long-standing challenge in the machine learning community. Training of deep neural networks on highly imbalanced labels may converge to local minima, resulting in suboptimal inference. There are still efforts underway to handle such class imbalance situation of isointense labels in the medical image domain. Hashemi et al. [36] proposes an exclusive multi-label multi-class training strategy for infant brain tissue segmentation. Also, Milletari el al. [37] uses Dice coefficient as an objective function for training of the FCN to address the imbalance between foreground and background. Furthermore, other similarity loss functions are introduced for detecting multiple sclerosis lesion in highly imbalanced data [38] and reducing the Hausdorff Distance in segmentation [39].

Creating high quality labels on clinical data requires human expertise and thus access to labeled data is very limited. There have been several strategies to deal with a limited number of labeled data in semi-supervised ways [40], [41]. Roy et al. [40] employs a popular brain segmentation tool to obtain auxiliary labels for pre-training. The pre-trained model is then fine-tuned on the limited manual labels. Radosavovic et al. [41] introduces a data distillation approach that generates extra labels by ensembling predictions from teacher models trained on different transformations of unlabeled data and then re-train a student model on the generated labels.

Our main contribution is threefold: 1) we introduce dilated dense blocks with exponentially increasing dilation rates to encode multi-scale contextual information without consecutive max-pooling and additional complexity in the encoding path. The new encoding path is integrated into a decoder of FC-DenseNet (hereafter, FC-Dense ContextNet). 2) We propose to independently learn label probabilities of dentate and interposed nuclei to handle a class imbalance problem with the multi-class hybrid asymmetric loss function. We incorporate an attention score map as a regularization term and an overlap penalty for avoiding overlaps between dentate and interposed nuclei into the total loss function. 3) We exploit self-training strategies to overcome the problem of limited labeled data, which allow the proposed network to utilize auxiliary labels for improving training. To the best of our knowledge, this is the first work to segment simultaneously deep cerebellar dentate and interposed nuclei using a deep neural network on the patient-specific MRI data.

The rest of the paper is organized as follows. We first detail aforementioned contributions in Section II. In Section III, we briefly describe the experimental setup. We then present and discuss segmentation results and carry out ablation study to demonstrate the effectiveness of the proposed components, followed by presenting limitations and future directions in Section IV. Finally, we conclude this work in Section V.

## II. METHODS

In this section, we first extend FC-DenseNet to effectively encode contextual information at different scales. Also, we explain how to mitigate the class imbalance problem of dentate and interposed nuclei in the network followed by presenting the proposed loss function. Finally, self-training strategies are presented for handling limited labeled data. To formulate the training problem, let us denote training set by $T = \{(\mathbb{U}_i, \mathbb{L}_i), i = 1, \ldots, n\}$, where $\mathbb{U}_i = \{u_{ij}, j = 1, \ldots, m\} \in \mathbb{R}^m$ is the input image patch and $\mathbb{L}_i = \{l_{ij}, j = 1, \ldots, m\} \in \mathbb{R}^m$, $l_{ij} \in c$ is the ground truth label patch. $n$ is the number of training samples, $m$ is the number of voxels in the image patch, and $c$ is a label class index. We use region of interest (ROI) images around the DCN from whole brain images for efficient training. During the inference, the ROI on the test image is efficiently localized using the anatomical similarity between training images and test image (see Fig. 2). Overlapping 3D patches (sub-volumes) on the ROI are utilized as input of the model due to memory constraints and a limited number of training data. Overview of our proposed DCN segmentation framework (DCN-Net) is presented in Fig. 3. Each proposed component is detailed next.

### A. Deep Context-Aware Feature Learning

FC-DenseNet [25] has been successfully applied in many segmentation tasks [25], [26], [38]. However, it still requires consecutive max-pooling to encode features on a larger receptive field, resulting in loss of detailed boundary information, especially that is critical in small structures.

A dilated convolution layer [31] adjusts the size of the receptive field by using a sparse convolutional kernel and thus can be exploited in the network to address the above problem without adding complexity. Inspired by [42], we propose to use dilated convolutional layers with exponentially increasing dilation rates in each dense block of the encoder (referred as to a dilated dense block). As shown in Fig. 3-(a), the number of convolutional layers (L) in the first, second, and third dilated dense block is 3, 4, and 5, respectively. Dilation rates (1, 1, 2), (1, 1, 2, 4), and (1, 1, 2, 4, 8) are applied to convolutional layers of three dilated dense blocks. Such growing dilation rates allow us to avoid the gridding effect [42] in cascaded dilated convolutions with the same rate. Moreover, each dilated dense block aggregates contextual information at different scales, similar to the atrous spatial pyramid pooling [32]. Also, the number of channels in convolutional layers of each dilated dense block grows by 8.

We replace transition down blocks and dense blocks in the encoding path of FC-DenseNet with only dilated dense blocks as illustrated in Fig. 4-(a). Also, in order to build a deeper network without a memory burden for 3D input patches, we add max-pooling operations in the skip-connections. Intermediate feature maps at different scales are then concatenated into the decoder path. This architecture enables the network to keep the rich semantic information while going deeper into the network. Finally, we incorporate multi-scale (pyramid) input patches into max-pooled feature maps in the skip-connection (see also Fig. 3-(a)). This strategy facilitates learning of locality aware features by recovering information lost by max-pooling, thereby further improving the segmentation performance [43]. We use the network described herein as a backbone (named FC-Dense ContextNet).

*B. Independent Label Probability Estimation*

Jointly learning representations of highly imbalanced class labels with similar intensity on the image might cause suboptimal label prediction since it relies mostly on prevalence labels (e.g., in this work the ratio for the average number of voxels of dentate and interposed nuclei is 12.6:1).

To reduce the bias in training with imbalanced dentate and interposed nuclei labels, we propose an independent single-label multi-class training strategy which separately learns dentate and interposed nuclei label probabilities in a single network (see Fig. 4-(b)). Specifically, label probabilities of background/dentate $p_{ijc}^D$ and background/interposed $p_{ijc}^I$, respectively, for the training sample $i$, the voxel $j$, and the label class index $c$ are independently estimated using softmax activation at the final layer of the network:

$$p_{ijc}^D = \mathrm{P}(l_{ij}^D = c\,|\mathbb{U}_i, \Theta),\ c \in \{1:\text{background},\ 2:\text{dentate}\} \text{ and } p_{ijc}^I = \mathrm{P}(l_{ij}^I = c\,|\mathbb{U}_i, \Theta),\ c \in \{1:$$

background, 2: interposed nuclei\}, (1)

where $l_{ij}^D$ is a ground truth background/dentate label and $l_{ij}^I$ is a ground truth background/interposed label for a voxel $j$, given the image patch $\mathbb{U}_i$ of a training sample $i$ and network weights $\Theta$. The ground truth label maps $\mathbb{L}_i^D$ and $\mathbb{L}_i^I$ are one-hot encoded:

$$g_{ijc}^D = \begin{cases} 1, & \text{if } l_{ij}^D = c \\ 0, & \text{others} \end{cases} \text{ and } g_{ijc}^I = \begin{cases} 1, & \text{if } l_{ij}^I = c \\ 0, & \text{others} \end{cases}. \tag{2}$$

The Dice coefficient loss has been widely used for medical image segmentation [37]. However,

since it equally weighs false positive and false negative, resulting in segmentation with low recall, it may not be suitable for highly imbalanced small objects [38]. In this study, we introduce a multi-class hybrid asymmetric loss which handles such a class imbalance problem of small objects by weighing false negative for higher recall and moreover focuses more on low probability classes (hard examples).

Tversky loss (*TL*) [44] generalizes the Dice coefficient loss by balancing false negative and false positive. Given estimated label probability map $p_{ijc}$ and one-hot coded ground label map $g_{ijc}$, the loss function is defined as:

$$TL_c = 1 - \frac{1}{n}\sum_i \frac{\sum_j p_{ijc} g_{ijc}}{\sum_j p_{ijc} g_{ijc} + \alpha \sum_j p_{ij1} g_{ij2} + \beta \sum_j p_{ij2} g_{ij1}}. \tag{3}$$

where $\alpha$ and $\beta$ control the influence of false positives and false negatives. In this work, we set $\alpha$ and $\beta$, respectively, as 0.3 and 0.7 to improve recall, weighing false negatives. This may lead to higher performance and generalization for segmentation of imbalanced small structures [44]. Also, Focal loss (*FL*) [45] extends the cross entropy loss by adding a coefficient $(g_{ijc} - p_{ijc})^2$ to attend to the lower probability class during the training:

$$FL_c = -\frac{1}{nm}\sum_i \sum_j (g_{ijc} - p_{ijc})^2 \log(p_{ijc}). \tag{4}$$

Finally, the proposed hybrid segmentation loss for dentate and interposed nuclei, respectively, combines multi-class *TL* and *FL*:

$$\mathcal{L}_D(\mathrm{T};\Theta) = \sum_c w_c (\pi_\mathrm{T} TL_c^D + \pi_\mathrm{F} FL_c^D) \text{ and } \mathcal{L}_I(\mathrm{T};\Theta) = \sum_c w_c (\pi_\mathrm{T} TL_c^I + \pi_\mathrm{F} FL_c^I). \tag{5}$$

The class weight $w_c$ is computed from a volume ratio of dentate or interposed nuclei and background. $\pi_\mathrm{T}$

and $\pi_F$ are weights for *TL* and *FL*, respectively, and equally set to 0.5.

While Hashemi et al. [36] uses a sigmoid function on cerebrospinal fluid and white matter labels, respectively, for whole brain tissue segmentation (multi-label multi-class problem), we estimate probabilities of dentate or interposed nuclei and background together using a softmax activation function (single-label multi-class problem).

*C. Attention Regularization and Overlap Penalty*

We incorporate an attention regularization term and overlap penalty into a total loss function to accelerate convergence by focusing more on relevant regions and further improve the label prediction without uncertainty on the border between dentate and interposed nuclei.

Motivated by Schlemper et al. [24] and Jetley et al. [46], we exploit an attention mechanism to effectively leverage the salient features in the network. We introduce an attention module which is incorporated into the skip-connection. As illustrated in Fig. 3-(b), the attention module calculates an attention score map to highlight meaningful regions and suppress feature responses irrelevant to segmentation. The number of channels of an intermediate feature map from the dilated dense block in the encoder is first changed to the number of channels of a gating signal (coarse feature) from the dense block in the decoder using 1×1×1 convolution layer. The gating signal is then up-sampled to the dimension of the feature map. The following channel-wise operations (average pooling, max pooling, and squeeze) involve in *where* to attend by learning the spatial dependency. Outputs of the channel-wise operations are concatenated followed by exponential linear transformation and 1×1×1 convolution layer. The attention score map is finally obtained using a sigmoid function. The attention module output both the attention score map and the intermediate feature map scaled by the attention score map.

The final attention score map ($a$) concatenates attention score maps from the attention module in the skip-connections followed by 1×1×1 convolution layer (Fig. 3-(a)). The softmax probability, given $a_i$ and $\Theta$ for the training sample $i$, the voxel $j$, and the class label index $c$ corresponds to:

$$p_{ijc}^A = \mathrm{P}\big(l_{ij}^{D+I} = c \mid a_i, \Theta\big), c \in \{1\text{: background, 2: dentate and interposed nuclei}\}, \quad (6)$$

where $l_{ij}^{D+I}$ is a ground truth background/dentate and interposed nuclei label. The ground truth label guides the final attention score map during the training by minimizing the attention loss as a regularization term which encourages attention to relevant features (Fig. 3-(a)). We utilize a categorical cross-entropy function for the attention loss:

$$\mathcal{L}_A(\mathrm{T}; \Theta) = -\frac{1}{nm}\sum_i \sum_c \sum_j g_{ij}^{D+I} \log\big(p_{ijc}^A\big), \text{ where the one-hot coded map } g_{ij}^{D+I} = \begin{cases} 1, & \text{if } l_{ij}^{D+I} = c \\ 0, & \text{others} \end{cases} \quad (7)$$

The independent label probability learning effectively mitigates a class imbalance problem in segmenting dentate and interposed nuclei. However, since dentate and interposed nuclei are in the same vicinity and independently estimated structures are not mutually exclusive, there might be oftentimes overlapping regions between segmented dentate and interposed nuclei. We therefore propose to impose an overlap penalty during training. To this end, we incorporate Dice coefficient (DC) [47] between estimated dentate and interposed nuclei labels as the overlap loss into a total loss function:

$$\mathcal{L}_O(\mathrm{T}; \Theta) = \frac{1}{n}\sum_i \frac{2\sum_j p_{ij2}^D p_{ij2}^I}{\sum_j (p_{ij2}^D + p_{ij2}^I)}. \quad (8)$$

Finally, the total loss function in our proposed DCN-Net can be defined as:

$$\mathcal{L}_{total}(\mathrm{T}; \Theta) = \lambda_D \mathcal{L}_D + \lambda_I \mathcal{L}_I + \lambda_A \mathcal{L}_A + \lambda_O \mathcal{L}_O, \quad (9)$$

where $\lambda_D, \lambda_I, \lambda_A,$ and $\lambda_O$ are weight values for the dentate and interposed nuclei segmentation losses, the attention loss, and the overlap loss. The weights are empirically set to 1.0, 1.0, 0.5, and 0.1, respectively, in this work.

*D. Semi-Supervised Learning*

Collecting a large amount of clinical data is crucial for effective training of deep learning models driven by supervised learning. Data augmentation is a popular way to increase training data by adding spatial variation of data [33]. Also, patch (or sub-volumes) based processing in medical imaging is typically considered to significantly increase the number of training samples, while reducing the memory burden [30]. Even though a vast amount of unlabeled data is given, creating high quality labels is a core task to train a model. However, it requires anatomical expertise and is labor-intensive.

Inspired by [40] and [41], we utilize two self-training strategies that use predictions on unlabeled data to refine a model, handling data limitation as shown in Fig. 5. Our database contains 29 labeled data (pairs of 7T diffusion-weighted B0 MRIs and corresponding manual labels) and 31 unlabeled 7T B0 MRIs.

First, given 31 unlabeled 7T B0 MRIs, corresponding dentate and interposed nuclei labels are predicted using initially trained DCN-Net on 29 manually labeled data. K pseudo labels are selected via visual inspection. The DCN-Net is pre-trained on the pseudo labeled data and is then fine-tuned on 29 manually labeled data (Fig. 5-(a)). While Roy et al. [40] uses a different automatic segmentation tool to obtain pseudo labels, we predict the labels using our own model trained with manually labeled data and use them to update the same model. The pre-training encourages the model to have a good initialization by leveraging pseudo labels from a wide variety of unlabeled data. Fine-tuning of the pre-trained model contributes to prune the discrepancy between a limited number of manual labels and pseudo labels.

We also exploit knowledge distillation [48] based on randomness of the model initialization as illustrated in Fig. 5-(b). We train the DCN-Net (randomly initialized) N-times on manual labels and then predict dentate and interposed nuclei labels on 31 unlabeled 7T B0 MRIs using N trained models. N predicted labels on each unlabeled data are fused via a majority voting to generate an auxiliary label (N = 5 in this work). We finally re-train the DCN-Net on a union set of 31 auxiliary labels and 29 manual labels. Ensembling predictions of transformed data in a single model can be more efficient than fusing predictions from multiple trained models [41], but in the medical image domain, since such data transformation might add biases on the shape of anatomical structures, we leverage the learned knowledge in randomly initialized models.

## III. EXPERIMENTAL SETUP

We performed extensive experiments to validate our proposed model in deep cerebellar dentate and interposed nuclei segmentation. In this section, we present datasets, pre-processing, details about implementation and training, and evaluation metrics used for validation.

### A. *Data and Pre-processing*

The 7T diffusion-weighted MRIs (B0) of 60 subjects were used in this work under approval of the Institutional Review Board at the University of Minnesota. Preprocessing steps included motion, susceptibility and eddy current distortions correction using FSL [49] and are detailed in [50]. The voxel size of the B0 image is $1.25 \times 1.25 \times 1.25 \text{mm}^3$.

29 B0 images were randomly chosen for validation. Of deep cerebellar nuclei, fastigial nuclei could not be manually segmented due to low contrast in the image. We thus segmented deep cerebellar dentate and interposed nuclei which are visible on the B0 image. Dentate and interposed nuclei were manually labeled on the images and carefully cross-validated by three anatomical experts, and were served as ground truth [51]. Remaining 31 B0 images where there do not exist ground truth labels were used to create extra labels for semi-supervised learning (as described in section II-D).

We set bounding box regions around the dentate and interposed nuclei labels on whole brain images to facilitate training. For the ROI localization on a test image during the inference, a reference training image is chosen by measuring a similarity score (e.g., mutual information) [52] between training images and the test image. The reference image is linearly registered onto the test image and corresponding dentate and interposed nuclei mask is then transformed onto the test image (see Fig. 2). This is considered more suitable than rather applying different architectures (e.g., Mask R-CNN [53]) to localize the bounding box around small target structures due to spatial similarity of brain anatomy.

*B. Implementation and Training Details*

We implemented the proposed DCN-Net using Keras [54] library package on top of Tensorflow [55]. We compared DCN-Net with the existing DCN segmentation tools (SUIT [11]) and publicly available state-of-the-art deep neural networks (DNNs) – DeepMedic [21], LiviaNet [22], U-Net [23], Attention U-Net [24], FC-DenseNet [25], DeepLab v3+ [35] (with FC-DenseNet as a backbone), and CE-Net [33]. To this end, we utilized the SUIT toolbox available in SPM12 [56] and adopted original implementations of the networks. 2D architectures are extended to 3D for volumetric segmentation of dentate and interposed nuclei.

For fair comparison, we trained models under the same environment. The networks were initialized using the approach proposed in [57]. Random seeds in kernels were fixed during the training to avoid biases from different initialization. The parameters were optimized with Adam [58] with 0.001 learning rate. The proposed multi-class hybrid asymmetric loss was applied in other networks to eliminate the effect of different loss functions in segmentation performance. The size of mini-batches is 8 and the number of epochs is 50. Training is early stopped after 20 epochs without improvement on a validation set and then we take model weights with a minimal validation loss to avoid overfitting on test set.

3D patches (sub-volumes) based processing within each network is considered to reduce the memory burden and to significantly increase the number of training samples. The size of input patches is $32 \times 32 \times 32$ (the effect of different patch sizes in segmentation is compared later in Section IV-D). Predicted output patches are the same size as input patches. The step size for patch extraction is $5 \times 5 \times 5$ (patch step size 9×9×9 is also used for investigating the effect of the number of training samples in segmentation (Section IV-C)). Training of networks was done with NVIDIA Tesla K40 GPU. The source code and relevant materials will be available from the corresponding author upon reasonable request.

*C. Evaluation Metrics*

DC [47], center of mass distance (CMD), mean surface distance (MSD) between ground truth and segmented results of each method, and volumes were computed for quantitative analysis. Those metrics are considered highly relevant to accurately represent any small anatomical structures [52].

For statistical analysis of each measure, a paired t-test was performed on single comparisons. A one-way analysis of variance (ANOVA) and Tukey's honest significance post-hoc test were conducted for multiple comparisons. Five-fold cross validation on 29 test sets was used for evaluation. 20% of the training data was used as validation set. Thus, we trained the models on five combinations of training and test data in which each combination consists of 19 or 20 training sets, 4 validation sets, and 5 or 6 test sets. Note that, since we have utilized overlapping 3D patches (e.g., 32x32x32 size and 5x5x5 step) as input of the models, the number of training samples (patch images) used in the model was approximately 3500 at each fold (which is discussed in Section IV-C).

## IV. RESULTS AND DISCUSSION

In this section, we present quantitative and qualitative segmentation results of our proposed model with comparison to the existing DCN segmentation tool and state-of-the-art deep learning architectures. We investigate the effect of the number of training samples in the segmentation performance of each deep learning model. The influence of the size of image patches in segmentation within the proposed DCN-Net is also explored. Moreover, we provide in-depth study on the impact of each component in our proposed model followed by discussing limitations and future direction.

*A. Quantitative Comparison with the Existing Tool*

Of cerebellum parcellation methods presented in [13], SUIT [11] is able to localize the DCN on subject-specific data using the probabilistic atlas of the cerebellum. SUIT also provides a tool specific to studying the dentate for a group of subjects but instead of segmenting the dentate, it aims at optimizing the normalization of the subject data to an atlas space by using user-provided dentate masks [17]. Since it is different from the purpose of this study, we compared segmentation results obtained by using SUIT (with DARTEL normalization [59]) with segmentation based on the proposed DCN-Net (without self-training strategies).

As shown in Fig. 6, DCN-Net significantly outperformed in every metric for dentate and interposed nuclei segmentation ($p<0.001$). The consistency of SUIT segmentation was also much worse. Such a large variance was attributed to uncertainty in registration processes that might be influenced by anatomical variability in different populations, discrepancy of magnetic field strength, contrast, or resolution between atlas template and subject-specific data, and robustness of normalization techniques [52]. This confirms that atlas-based segmentation requires additional refinement steps. Further, the inference time of SUIT (~30min) was much longer than that of DCN-Net (~0.5min).

*B. Quantitative Comparison with State-of-the-art DNNs*

Dentate and interposed nuclei segmentation results obtained by using U-Net [23], Attention U-Net [24], FC-DenseNet [25], DeepLab v3+ [35], CE-Net [33], and DCN-Net (without self-training strategies) are quantitatively compared in Fig. 7. Note that training DeepMedic [21] and LiviaNet [22] failed to converge due to local minima in spite of using the proposed multi-class hybrid asymmetric loss.

DCN-Net produced significantly better dentate segmentation results than any of the state-of-the-arts ($p<0.05$ in MSD and DC). In interposed segmentation, DCN-Net also showed better performance than state-of-the-art methods in terms of average errors, which was mostly not statistically significant ($p>0.05$).

Overall, average errors and deviation in interposed segmentation were larger than those in dentate segmentation. This might be caused by the smaller size of interposed nuclei [60]. On the other hand, volume of interposed segmentation was closer to the manual label than dentate segmentation. Volume difference between dentate segmentations and the manual label was significant ($p<0.05$).

Table I gives the number of trainable parameters in each network. Densely connected network based models - FC-DenseNet, DeepLab v3+, and DCN-Net - have much fewer parameters due to their unique architecture that does not require the re-learning of redundant feature maps, that is important to reduce overfitting in training, resulting in better segmentation [25]. Note that DCN-Net has a comparable number of parameters with FC-DenseNet while it maintains the size of input patches without max-pooling in the encoder due to proposed dilated dense blocks, which allows for capturing contextual information

*C. The Effect of the Number of Training Samples*

To investigate the effect of the number of training samples in segmentation, we compared segmentation results obtained by using models trained with patches of $9\times9\times9$ (e.g., 765 training samples) and $5\times5\times5$ step sizes (e.g., 3,453 training samples). As shown in Fig. 7, training models with patches of $9\times9\times9$ step size worsened segmentation performance. The larger patch step size (i.e., less training samples) influenced more dentate segmentation than interposed segmentation. Particularly, while attention U-Net, FC-DenseNet, DeepLab v3+, and CE-Net with less training samples yielded significantly worse dentate segmentation ($p<0.05$ in DC), DCN-Net and U-Net produced comparable results with both patch step sizes.

Deterioration in interposed segmentation obtained by using networks with less training samples was not statistically significant ($p>0.05$). DCN-Net still achieved significantly better performance in dentate segmentation than state-of-the-art methods with less training samples ($p<0.05$ in MSD and DC). In interposed segmentation, DCN-Net also produced lower average error than state-of-the-art networks.

Particularly, DCN-Net significantly outperformed U-Net and CE-Net ($p<0.05$ in MSD and DC) and FC-DenseNet ($p<0.05$ in DC).

We observed that the performance of DCN-Net trained with less training samples is comparable to state-of-the-art models trained with more training samples both in dentate and interposed nuclei segmentation. This might prove the robustness of DCN-Net to the smaller number of training samples.

*D. The Effect of the 3D Patch Size*

We conducted an additional experiment to investigate the effect of different image patch sizes in segmentation of dentate and interposed nuclei. Patch-based processing is preferred to significantly increase training samples and handle memory issues, especially for volumetric medical image segmentation. For example, there are only 29 training images in our case, but 3,453 training samples (on the ROI images) are available when using $32\times32\times32$ patch size and $5\times5\times5$ step size. Since the larger patches may cause memory burden during training while the smaller patches are insufficient to encode contextual information, it is thus important to select a proper image patch size in the network.

Segmentation results obtained by using our proposed DCN-Net trained with $16\times16\times16$, $24\times24\times24$, $32\times32\times32$, and $40\times40\times40$ patch size with $5\times5\times5$ step size were quantitatively compared in Fig. 8. Using the larger patch sizes ($32\times32\times32$ and $40\times40\times40$) produced significantly better dentate segmentation than using the smaller patch size ($p<0.05$ in DC). This might explain that the larger patch size allows the model to encode the contextual information on the dentate region. For interposed nuclei segmentation, using the patch size of $32\times32\times32$ showed slightly better DC value than using other patch sizes. Interestingly, the patch size of $40\times40\times40$ showed the worst DC value and thus it might not be encouraged for encoding features on such smaller region. Further, it resulted in high computational overhead so that training took much longer time than using smaller patch sizes.

The patch size of $32\times32\times32$ is considered an optimal choice for better dentate and interposed nuclei segmentation (in terms of DC value) in this work.

*E. Qualitative Analysis*

Figures 9 and 10 visualize dentate and interposed nuclei segmentation on the 7T B0 MRI of a specific subject (see also a visual example in Fig. 1).

SUIT produced over-segmentation around the boundary of dentate and interposed nuclei. An incorrect registration between a low contrast test image and atlas template in SUIT might cause such a large systematic error in segmentation. State-of-the-art DNNs yielded better segmentation results than SUIT, but some artifacts were still observed on the region with similar intensity level, especially in the superior part (see the axial view of each method in Figures 9 and 10). Overall, segmentation results obtained by using DCN-Net were visually and quantitatively closer to the ground truth than segmentation based on state-of-the-art DNNs. This also exemplifies the robustness of our proposed DCN-Net to a low contrast image.

*F. Ablation Study*

We carried out an in-depth study on the impact of each component in DCN-Net for dentate and interposed nuclei segmentation. To this end, we performed segmentation based on DenseNet with the proposed dilated dense blocks (Dilated DenseNet), FC-DenseNet, FC-Dense ContextNet, and DCN-Net with and without self-training strategies. Segmentation results are quantitatively compared in Fig. 11. The effectiveness of proposed components is further studied next.

**1) Dilated dense blocks in the encoder and decoder style:** FC-Dense ContextNet (backbone) shares dilated dense blocks in the encoder with Dilated DenseNet and has the decoder to recover high-resolution features as FC-DenseNet does. FC-Dense ContextNet significantly outperformed Dilated DenseNet in dentate segmentation ($p<0.05$ in CMD and $p<0.001$ in MSD and DC) and also yielded significantly better interposed segmentation results ($p<0.05$ in DC), proving the effectiveness of the decoder in segmentation.

Moreover, FC-Dense ContextNet yielded significantly better dentate segmentation than FC-DenseNet ($p<0.05$ in DC) and also showed slightly more accurate interposed segmentation by exploiting dilated dense blocks in the encoder. To intuitively explain the effect of the proposed dilated dense blocks in FC-Dense ContextNet, we provide intermediate feature maps at each scale obtained from encoders of FC-Dense ContextNet and FC-DenseNet, respectively, using a 2D image example in Fig. 12. Compared with the dentate label of the input image, dilated dense blocks followed by max-pooling in FC-Dense ContextNet produced clearer feature maps than max-pooling followed by dense blocks in FC-DenseNet (see F2 and F3 in Fig. 12). This exemplifies the benefit of the dilated dense blocks in the FC-Dense ContextNet to extract features without max pooling operation followed by convolutional layers. Feature maps of interposed nuclei in both networks, however, were not clear due to its small size (appeared in a few voxels). To address this, multi-scale (pyramid) input patches are incorporated into feature maps after max-pooling in the skip-connection of the proposed DCN-Net and loss of details in interposed nuclei can be thus recovered in the decoder (see also the deep context-aware feature encoding block in Fig. 3-(a)).

In this study, we can clearly see that dilated dense blocks boost the feature representation power in the network, especially with higher impact of a decoder on segmentation.

**2) Multi-class hybrid asymmetric loss function**: Training the backbone network (FC-Dense ContextNet) with multi-class DC loss did not converge, while training the network with the proposed multi-class hybrid asymmetric loss converged very fast (See Fig. S-1 of the supplementary material for training loss and validation loss curve for the number of epochs). Unfortunately, the trained model with the DC loss failed to produce dentate and interposed nuclei segmentation results. This suggests that the DC loss may not be proper for segmentation of highly imbalanced small organs despite its popularity in medical image segmentation. The proposed multi-class hybrid asymmetric loss, however, was able to handle such a class imbalance problem in segmentation by adjusting weights of false negative and false positive for achieving higher recall, encouraging training to focus more on hard examples (low probability classes).

**3) Independent label probability estimation:** Independently estimating label probabilities yielded better segmentation results than jointly estimating label probabilities in FC Dense ContextNet in terms of accuracy and consistency as summarized in Table II. We observed that it is more effective in dentate segmentation (13.3% in CMD and 2.4% in DC) with the statistical significance ($p<0.001$ in MSD and DC) than interposed segmentation (2.3% in CMD and 0.5% in DC). This justifies that this strategy helps the network to avoid biases, especially in the larger label, induced by inter-dependency of highly imbalanced class labels during the multi-label probability estimation.

**4) Attention regularization and overlap penalty:** We investigated the effectiveness of attention and overlap losses in FC-Dense ContextNet with independent label probability estimation.

Table III summarizes each measure in dentate and interposed nuclei segmentation. While the attention regularization term slightly improved interposed segmentation, it has not been effective for dentate segmentation. This might explain that the attention loss helps detection of smaller organs, suppressing irrelevant features.

Fig. 13 visually compares dentate and interposed nuclei segmentation with and without the overlap loss to validate its effectiveness. We observed that the overlap penalty in the loss function reduces overlaps between dentate and interposed nuclei that might be caused by separately estimating label probabilities.

Both attention map and overlap penalty in the proposed FC-Dense ContextNet (DCN-Net) slightly improved interposed segmentation (2.7% in CMD and 1.5% in DC) while reducing overlaps between dentate and interposed nuclei. An interesting observation is that such an overlap constraint compensated for a slight deterioration in dentate segmentation obtained by FC-Dense ContextNet with only attention loss (see also volume difference between dentate segmentations when adding overlap loss in Table III).

**5) Self-training strategies:** We validated two self-training strategies described in Fig. 5 to effectively train the proposed DCN-Net on a limited number of labeled data. As shown in Table IV, while self-training strategies slightly improved interposed segmentation, it has not been effective for dentate segmentation. Similar to the impact of the attention loss, it might be difficult to further improve segmentation as it reaches theoretically optimal performance bound simulated by its volume [60]. For interposed segmentation, expanding the pool of training data (auxiliary and manual labels) via model distillation has shown to be more effective than pre-training with predicted labels and fine-tune training with manual labels (e.g., 0.701±0.16 and 0.686±0.14, respectively, in DC). Interestingly, pre-training with pseudo dentate and interposed nuclei labels (segmented by using DCN-Net trained on manual labels) achieved comparable results to DCN-Net trained on only manual labels. This proves that the pseudo labels were well segmented with DCN-Net and thus this facilitated semi-supervised learning.

Additionally, we trained U-Net with self-training strategies and performed dentate and interposed nuclei segmentation to validate the effectiveness of self-training in a different network. As summarized in Table V, self-training strategies significantly improved interposed segmentation performance of U-Net ($p<0.05$ in DC), that was thus comparable to that of DCN-Net with and without self-training strategies, while they have deteriorated dentate segmentation. Unlike pre-training of DCN-Net, pseudo dentate labels (obtained by using U-Net trained on manual labels) have impeded to pre-training of U-Net. Dentate segmentation obtained by using the pre-trained U-Net was significantly worse than dentate segmentation results from U-Net trained on only manual labels ($p<0.05$ in CMD, MSD, and DC). This indicates that the quality of pseudo dentate labels was not good enough for self-training and also explains deterioration in dentate segmentation obtained by using self-trained U-Net. Interposed segmentation obtained by using the pre-trained U-Net on pseudo labels, on the contrary, was improved and slightly better than interposed segmentation obtained by using pre-trained DCN-Net on pseudo labels (0.688±0.17 vs. 0.676±0.15 in DC). We demonstrate that high quality labels segmented by using DCN-Net trained on manual labels facilitate self-training of DCN-Net, thereby further improving segmentation results, especially in interposed nuclei.

*G. Limitation and Future Works*

We have validated DCN-Net only in our own dataset obtained from unique acquisition protocols since public datasets with DCN ground truth labels are not currently available. To evaluate generalizability and robustness of DCN-Net, unseen cases of a large scale data across centers should be explored and segmentation results need to be comprehensively analyzed.

Dentate and interposed nuclei have been segmented in this study using the B0 contrast from DWI and appear as hypo-intense regions. DWI is becoming part of the standard-of-care for imaging protocols in the clinical practice. Further, tractography based on DWI may allow for studying connections of the whole cerebellar system [61]. As demonstrated in many literatures [4], [9]–[11], [17], [62], various structural MR imaging modalities allow us to directly identify the DCN. Unfortunately, most of the modality 7T images in our database (e.g., susceptibility-weighted imaging and $T_2$-weighted image) were optimized to cover the basal ganglia region that was needed for deep brain stimulation studies and thus did not contain the cerebellum. Recently introduced quantitative susceptibility mapping (QSM) also exhibits clear visualization of sub-cortical brain structures due to higher levels of brain iron that induce susceptibility contrast [63]. However, 1) QSM is not part of the standard clinical imaging protocols and 2) it requires additional post-processing steps with uncertainty in susceptibility estimates under different acquisition protocols that still needs to be extensively investigated for its clinical use [64].

While we have considered the B0 diffusion MRI to visualize and segment the DCN in this study for the reasons, it should be noted that the proposed model can be exploited to segment the DCN on other clinically available MRI modalities where the nuclei are fairly visible due to its feature representation power, given sufficient training data. Furthermore, to find more suitable MRI contrasts for segmentation, the effectiveness of such MRI modalities in segmentation within DCN-Net can be evaluated. Also, learning an optimal combination of different contrasts, leveraging complementary information on the images should improve segmentation.

7T MR imaging enables direct identification of many anatomical structures thanks to its superior contrast and signal-to-noise ratio (SNR). Recently, the clinical use of 7T MRI has been approved by FDA. However, there are still a limited number of 7T MRI machines in current practice due to the significant hardware cost. It may therefore be needed to segment the DCN on standard clinical 3T MRI. Unfortunately, standard clinical MRI systems do not provide clear visualization of the DCN due to the poor SNR and contrast and thus oftentimes requires post-processing steps for enhancement (e.g., QSM). The quality of the manual segmentation for the ground truth labels on clinical 3T MRIs may not be sufficient for training or evaluation of the model. Future direction would be to utilize the 7T knowledge learned within the DCN-Net that can be transferred to segment the DCN on 3T MRIs (similar to 7T guided 3T MRI segmentation [52]).

The scalability of the proposed DCN-Net could allow for its wide use in practice. However, it was not fully assessed in this work due to the limited access to large-scale experts-labeled data. It is thus crucial to collect and label a large amount of MRI data with different magnetic fields, contrasts, and pathologies across various centers as discussed above. As more 7T systems are coming on-line and more labeled data will become available, the scalability of the proposed method on such large-scale datasets could be tested in an adequate manner.

Although the proposed DCN-Net was specialized in segmenting dentate and interposed nuclei, the issues handled in this paper - contextual information, class imbalance of small structures, and limited labeled data - are highly relevant to a slew of other applications in medical imaging. Further validation of DCN-Net on publicly available datasets to demonstrate its applicability to different segmentation tasks (e.g., brain tissue or subcortical structures) remains as a future work.

Finally, this work may allow neurosurgeons and clinicians to facilitate neuroanatomical studies of the DCN and/or dentate nucleus DBS planning by providing fast, accurate, and robust patient-specific deep cerebellar dentate and interposed nuclei segmentation. Particularly, dentate nucleus DBS treatment has been effective for post-stroke motor recovery [6]–[8]. Precise DBS lead placement within such a small nucleus is critical for maximizing benefits and minimizing side effects in the treatment [8]. A patient-specific volumetric dentate model provided by the proposed method may support the correct localization of the DBS lead. Moreover, such capabilities might lead to development of a simulation method for automatic DBS parameter optimization. The clinical feasibility of the proposed model in dentate nucleus DBS treatment needs to be further proven by retrospectively evaluating the post-operative electrode active contact locations on the dentate nucleus segmentation produced by our model.

## V. CONCLUSION

Volumetric segmentation of deep cerebellar nuclei is a pre-requisite for functional and neuroanatomical studies of the cerebellum. In this study, we proposed a novel fully convolutional deep learning architecture named DCN-Net for fast, accurate, and robust patient-specific segmentation of dentate and interposed nuclei. We introduced dilated dense blocks whose convolution layers have exponentially growing dilation rates so that the proposed network effectively encodes contextual information on different receptive fields. Also, we handled a class imbalance problem by independently estimating probabilities of dentate and interposed nuclei label. Moreover, self-training strategies were applied to facilitate the training of the proposed model by distilling data. Experimental results demonstrate that DCN-Net outperforms the existing atlas-based cerebellum segmentation tool and state-of-the-art deep neural networks in terms of accuracy and consistency even on lower contrast images. Also, in-depth analysis for the proposed components further validates the effectiveness of DCN-Net in dentate and interposed nuclei segmentation.

**Acknowledgements**

This work was supported in part by R01-NS081118, R01-NS113746, P50-NS098573, P30-NS076408 and P41-EB027061.

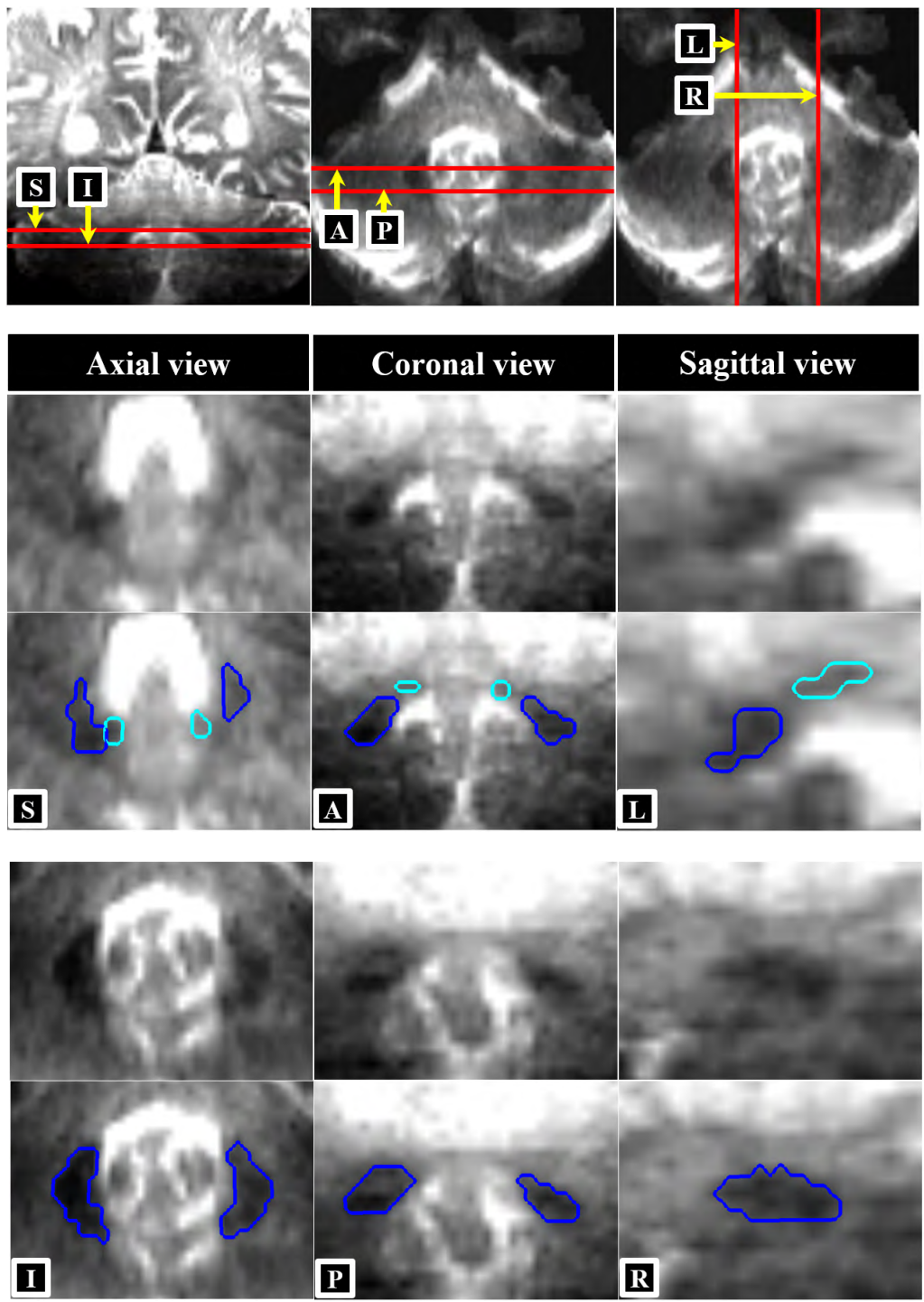

Fig. 1. Dentate and interposed nuclei hypo-intense region in selected planes of axial, coronal, and sagittal views on the 7T B0 MRI of a specific subject (top: two selected planes (red) in the whole brain image, middle: ROI and dentate (blue)/interposed (light blue) contours on superior (S), anterior (A), and left (L) planes of corresponding views, bottom: ROI and dentate (blue) contours on inferior (I), posterior (P), and right (R) planes).

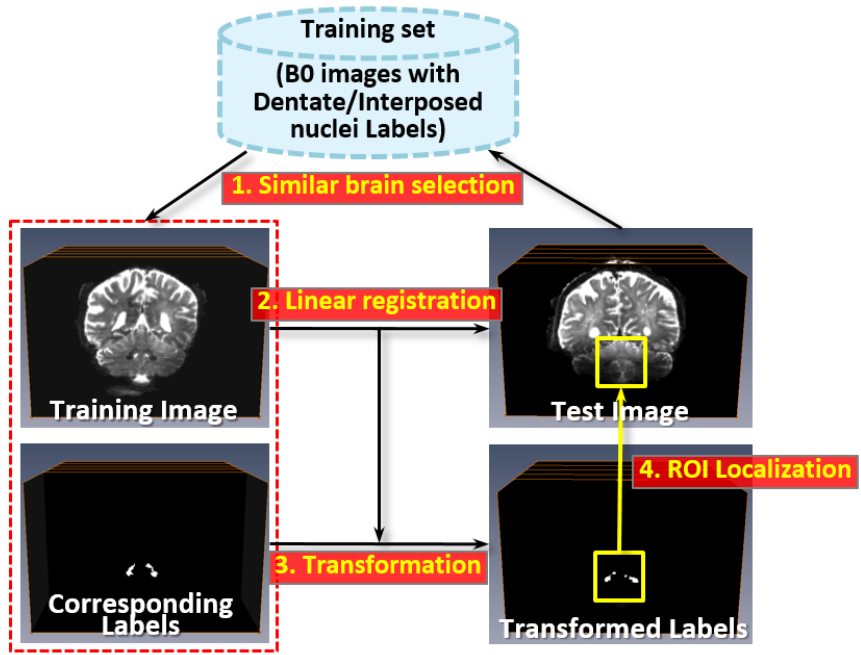

Fig. 2. Efficient ROI localization based on the anatomical similarity during the inference.

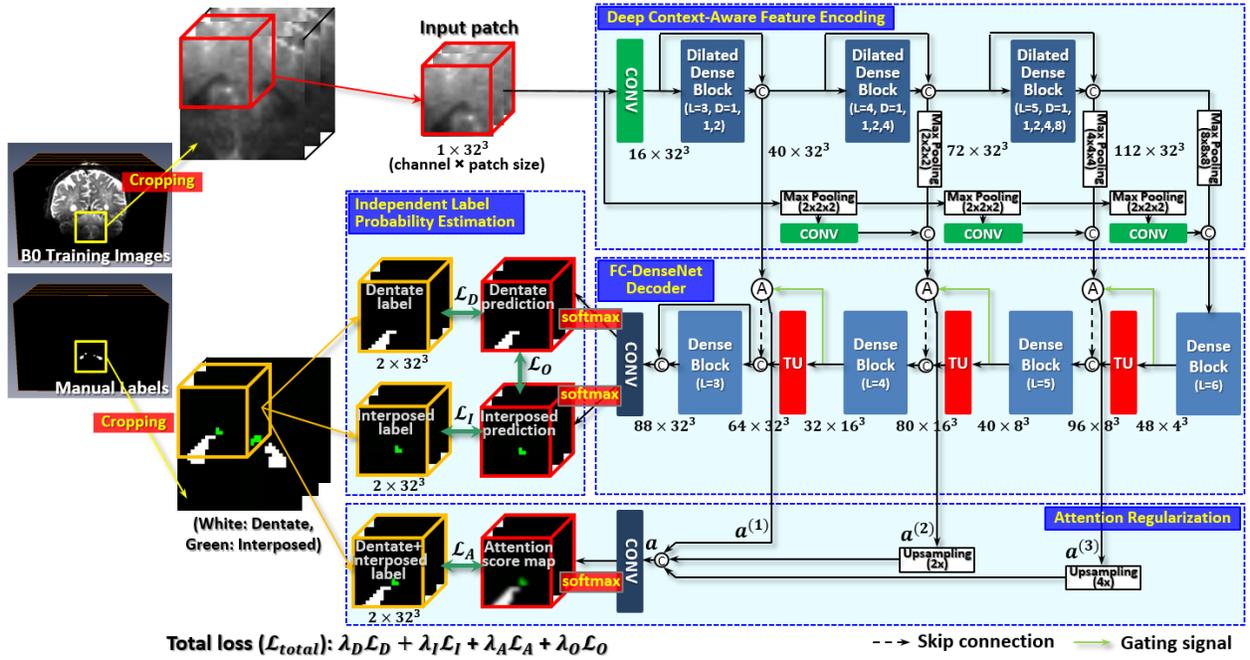

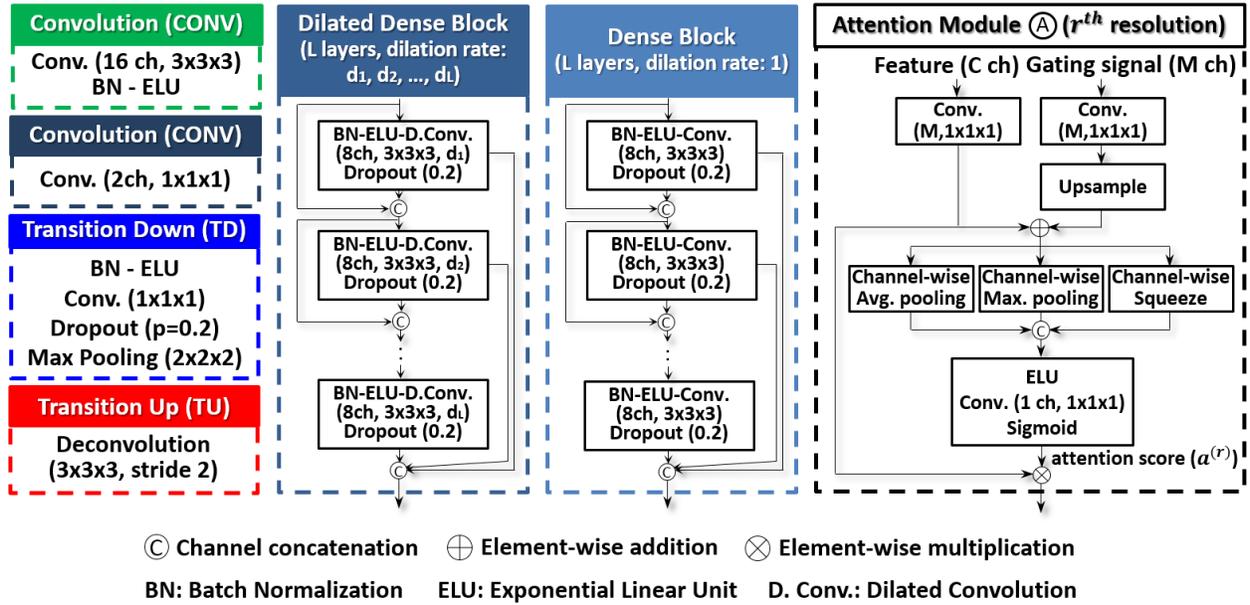

Fig. 3. Overview of the proposed framework for dentate and interposed nuclei segmentation. (a) Overall scheme of DCN-Net. It consists of deep context-aware feature encoder, decoder of FC-DenseNet, independent label probability estimation, and attention regularization. (b) Description of convolution layers, transition layers, dense block, dilated dense block, and the proposed attention module.

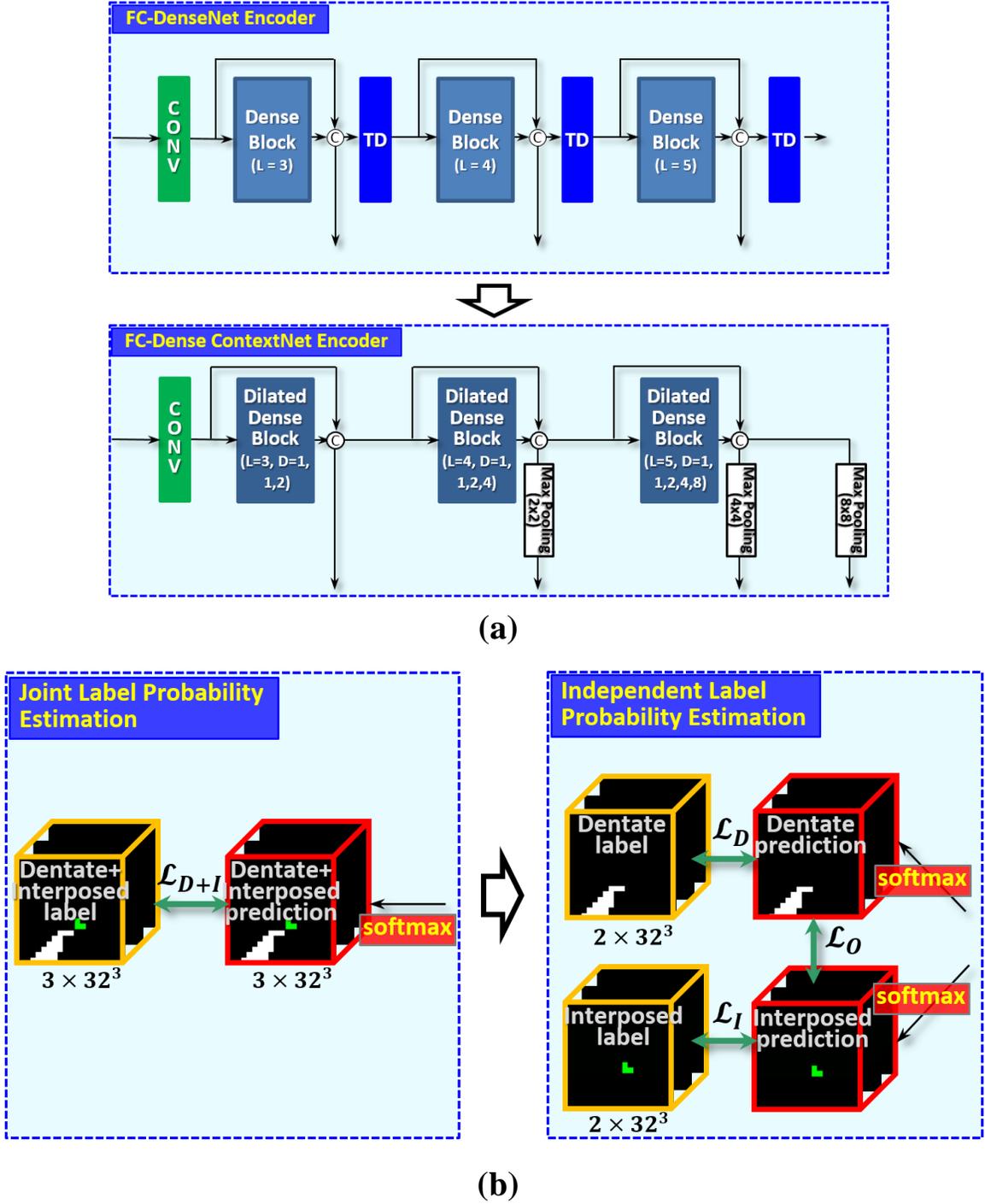

Fig. 4. Comparison of proposed architectures with existing models. (a) An encoder of FC-Dense ContextNet (with the proposed dilated dense blocks) replaces the existing encoder of FC-DenseNet for deep context-aware learning. (b) An independent label probability estimation with segmentation losses ($\mathcal{L}_D$ and $\mathcal{L}_I$) for dentate and interposed nuclei, respectively, and an overlap loss ($\mathcal{L}_O$) is proposed to handle multi-label dependency during the training via the existing joint label probability estimation with a multi-class segmentation loss for background, dentate, and interposed nuclei ($\mathcal{L}_{D+I}$).

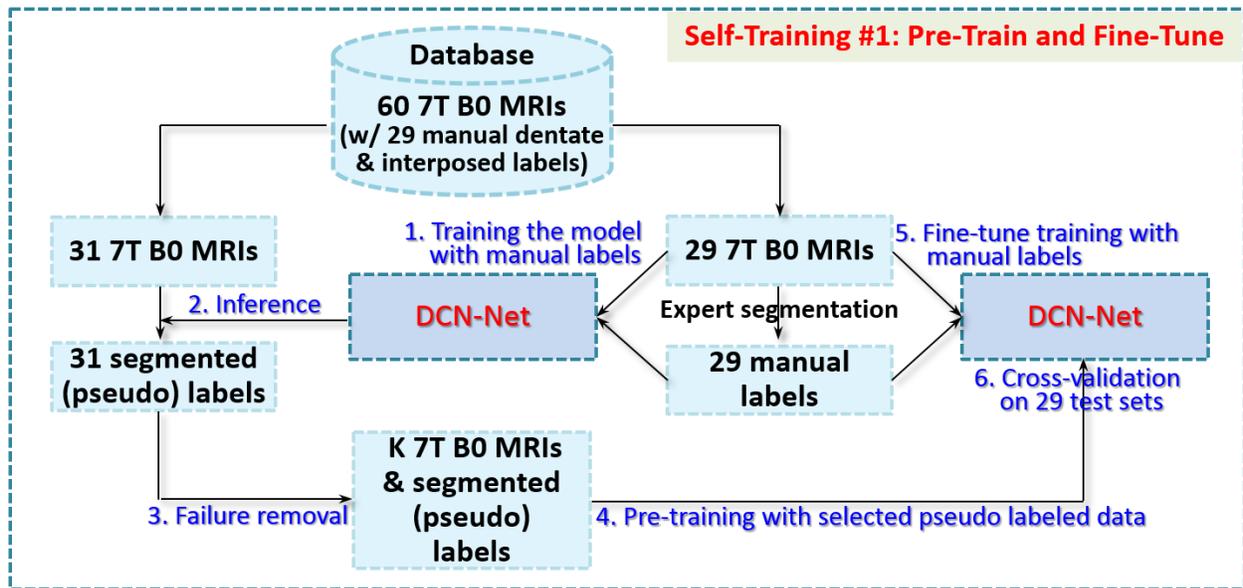

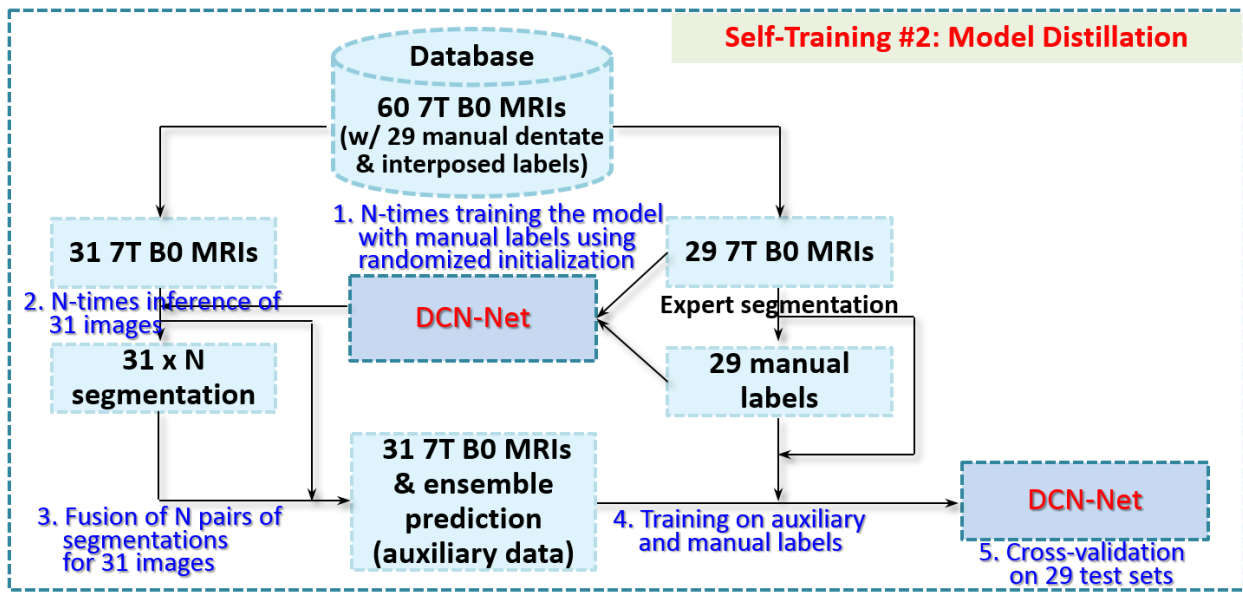

Fig. 5. The self-training strategies by (a) pre-training on predicted labels and fine-tuning on manual labels and using (b) an expanded pool of training data with ensemble of labels created by model distillation. K = 31 (without failures) and N = 5 in this work.

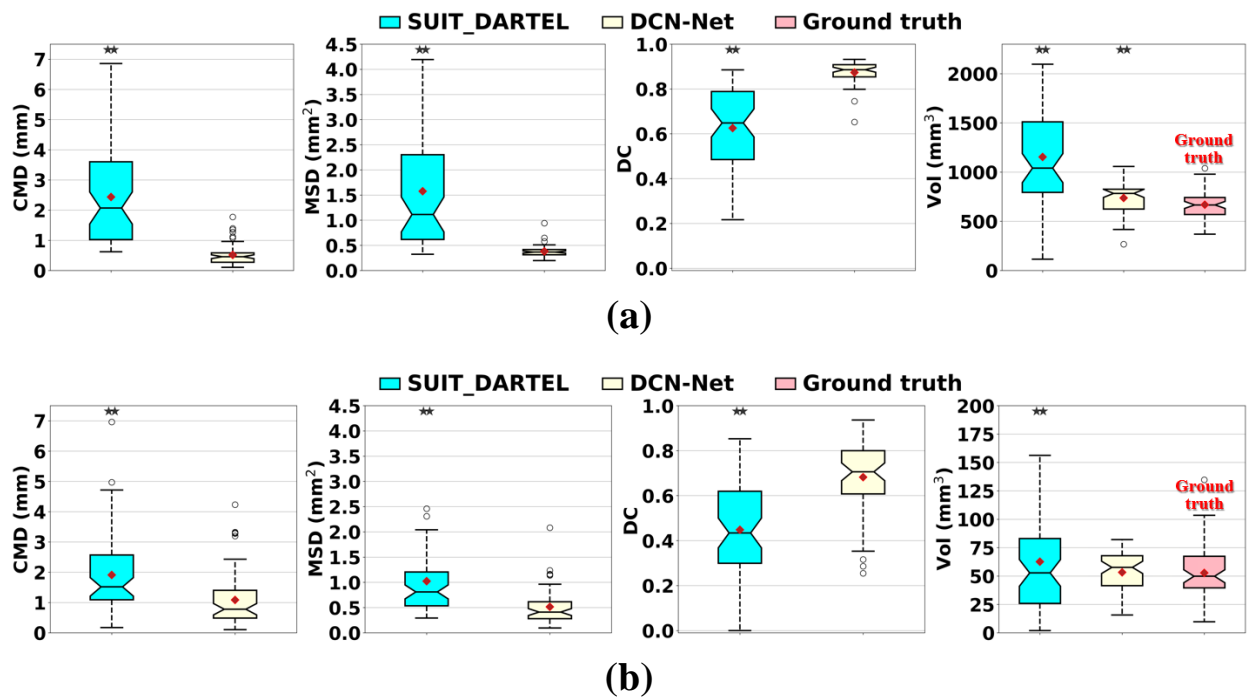

Fig. 6. Box plots of (a) dentate and (b) interposed nuclei segmentation based on SUIT and DCN-Net. *and **, respectively, indicate p<0.05 and p<0.001. See Table S-I of the supplementary material for average values and deviation.

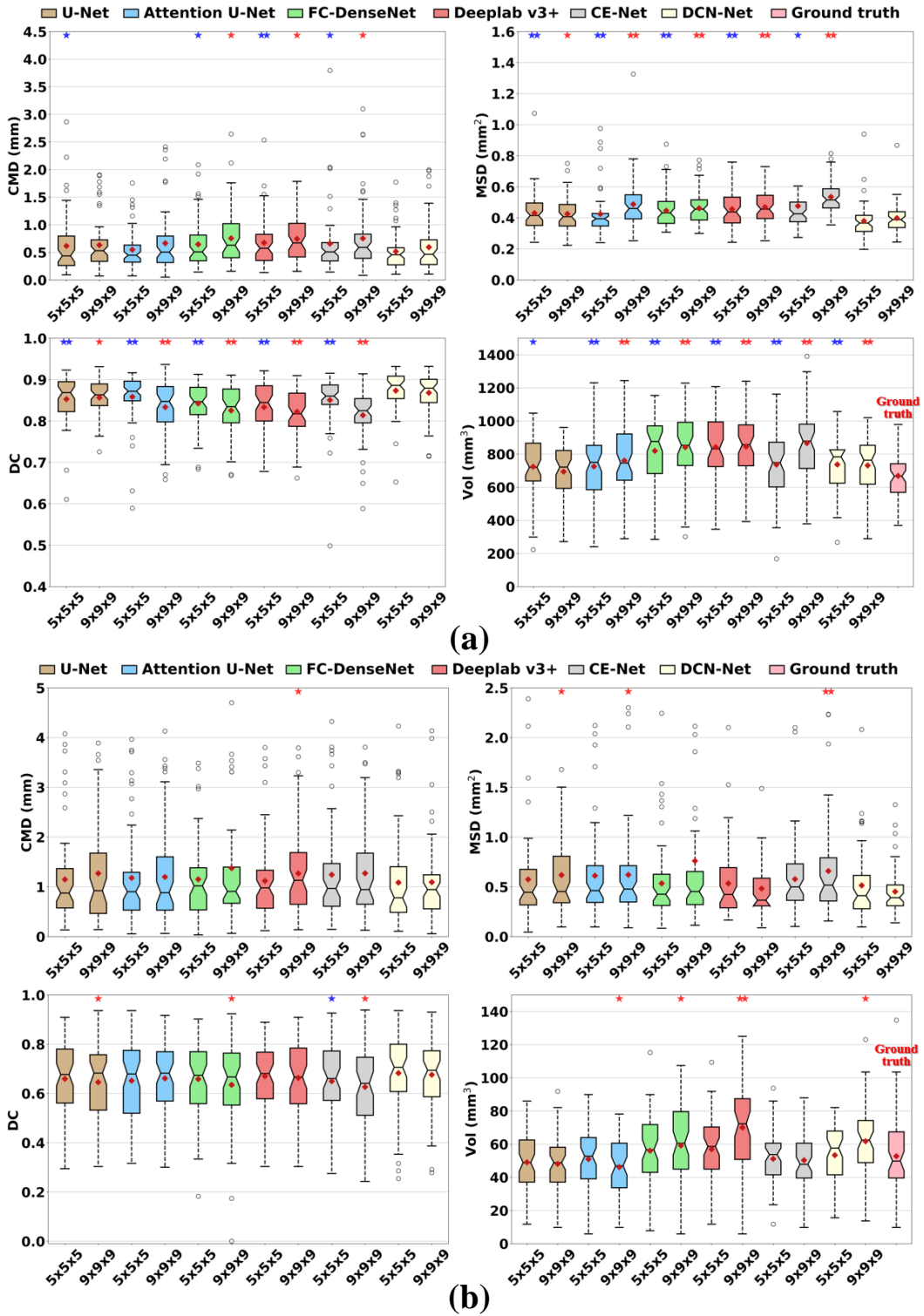

Fig. 7. Box plots of (a) dentate and (b) interposed nuclei segmentation based on state-of-the-art DNNs and DCN-Net with different patch step sizes. Statistical significance is marked as * and **, respectively, for p<0.05 and p<0.001 (blue for 5×5×5 patch step and red for 9×9×9 patch step). See Table S-II and S-III of the supplementary material for average values and deviation.

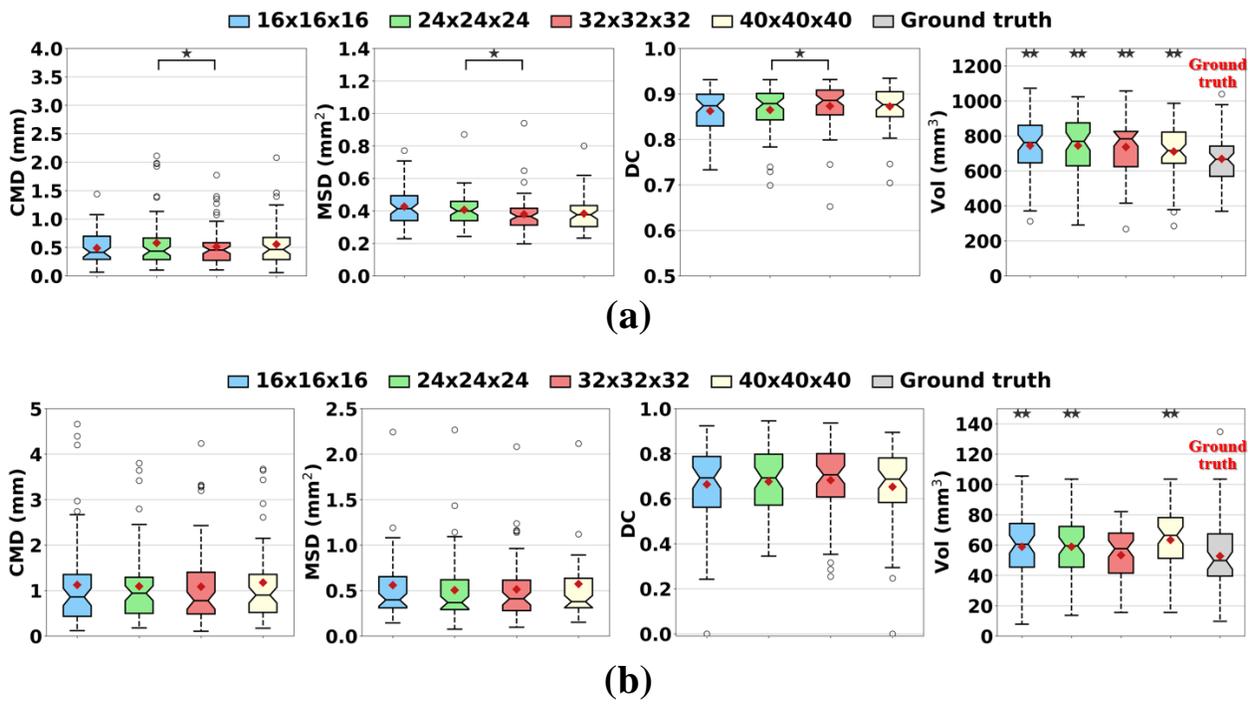

Fig. 8. Box plots of (a) dentate and (b) interposed nuclei segmentation based on DCN-Net with different patch sizes. Statistical significance is marked as * and **, respectively, for $p<0.05$ and $p<0.001$. See Table S-IV of the supplementary material for average values and deviation.

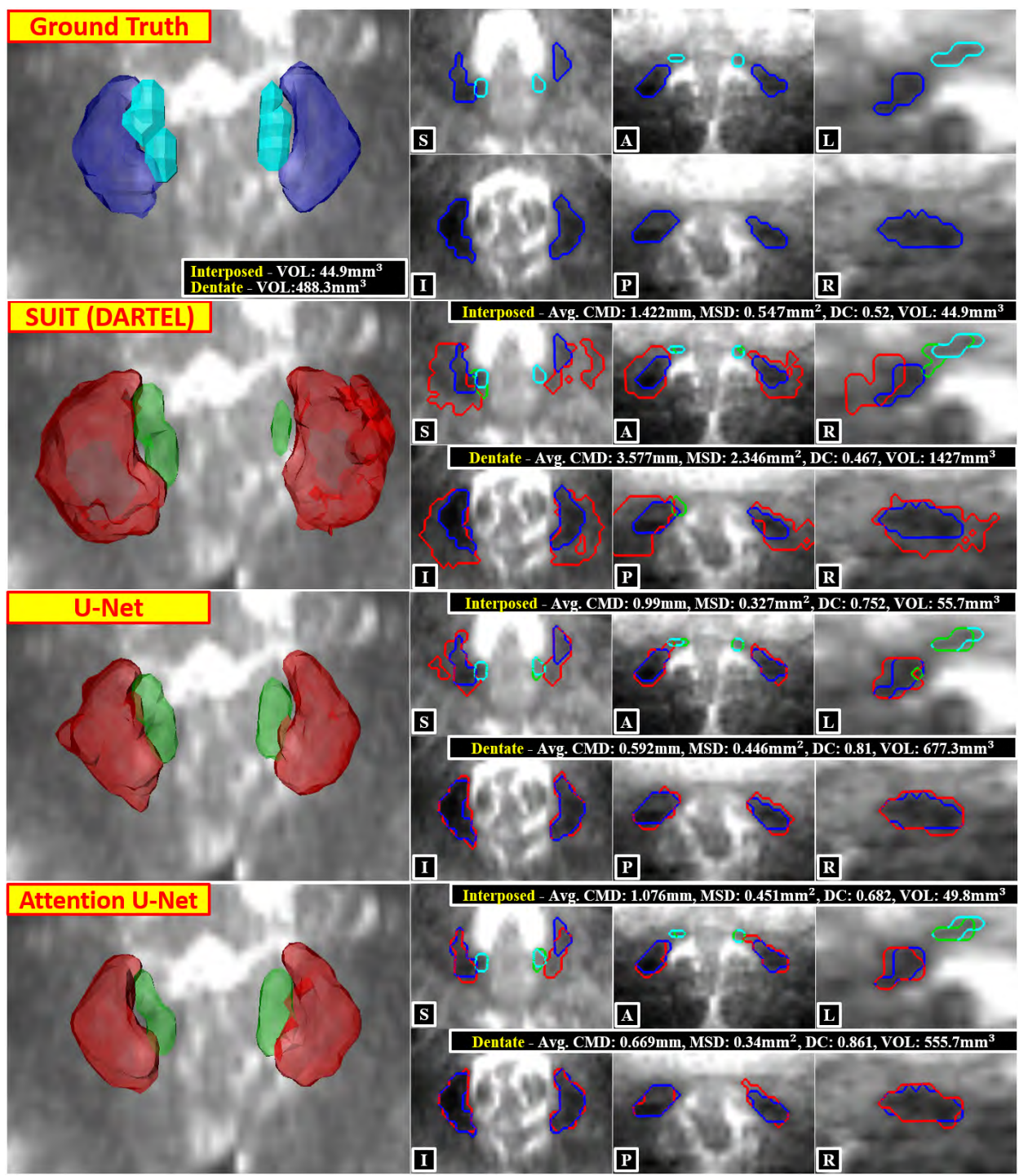

Fig. 9. Visual comparison of dentate and interposed nuclei segmentation results on a low contrast 7T B0 MRI of a specific subject. The first column shows ground truth dentate and interposed nuclei and volumetric segmentations obtained by SUIT (DARTEL), U-Net, and Attention U-Net. The last three columns are corresponding contours on two selected planes of Fig. 1 along with measures (average CMD, MSD, DC, and volume (VOL) in both sides). Red is the segmented dentate, green is the segmented interposed, blue is the ground truth dentate, and light blue is the ground truth interposed.

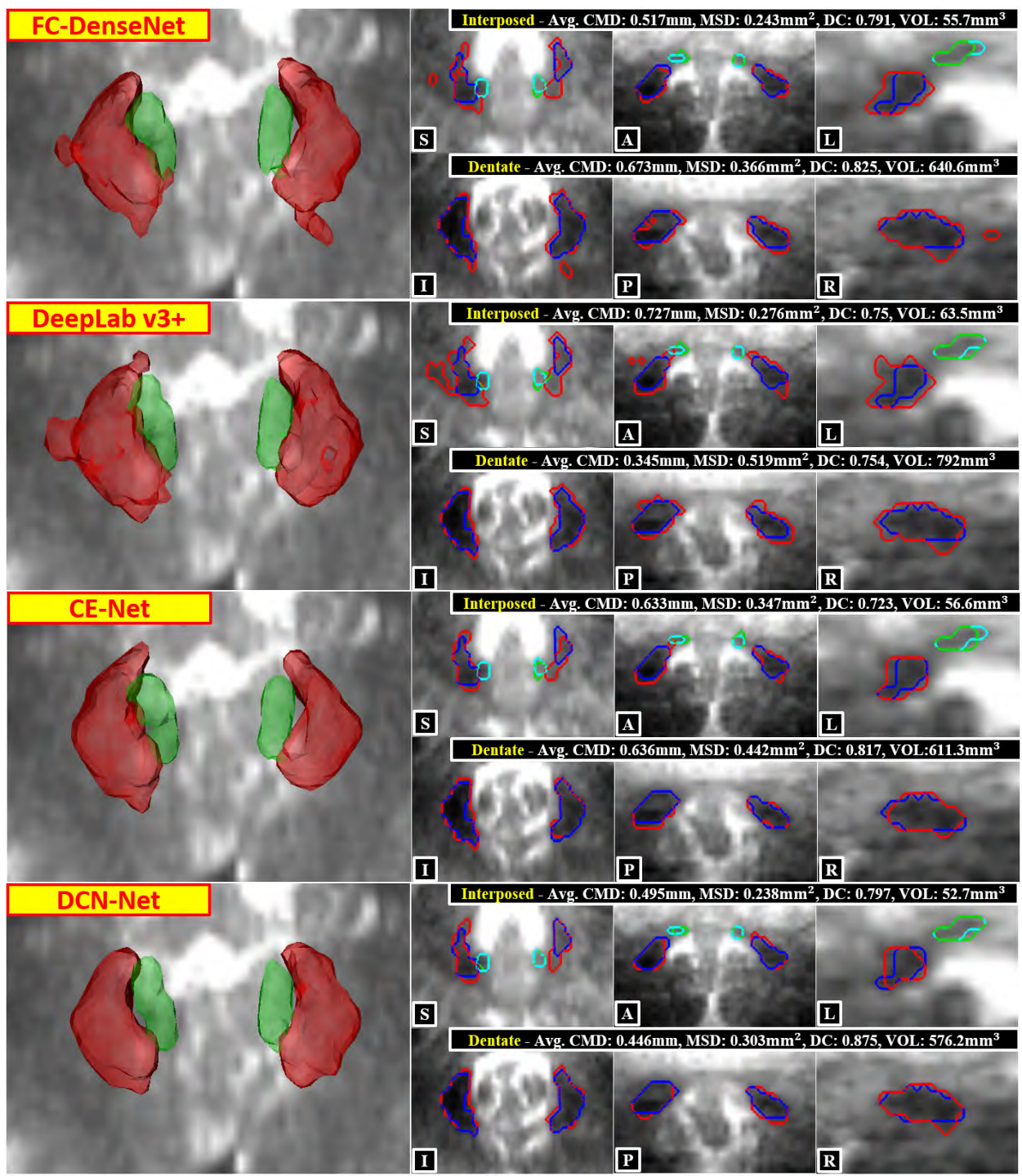

Fig. 10. Visual comparison of dentate and interposed nuclei segmentation results on a low contrast 7T B0 MRI of a specific subject. The first column shows dentate and interposed nuclei volumetric segmentations obtained by FC-DenseNet, DeepLab v3+, CE-Net, and the proposed DCN-Net. The last three columns are corresponding contours on two selected planes of Fig. 1 along with measures (average CMD, MSD, DC, and volume (VOL) in both sides). Red is the segmented dentate, green is the segmented interposed, blue is the ground truth dentate, and light blue is the ground truth interposed.

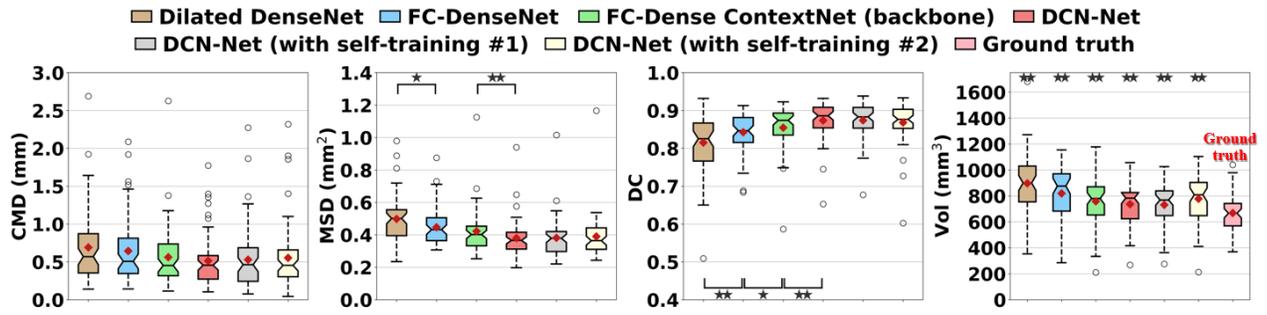

(a)

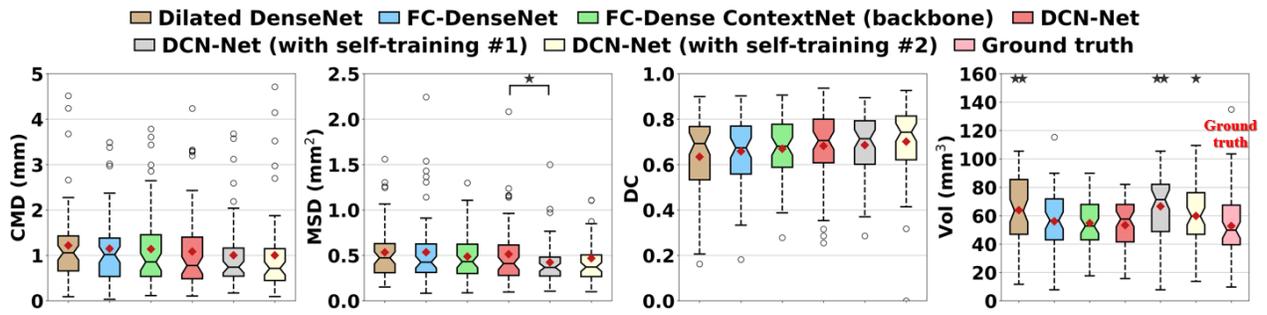

(b)

Fig. 11. Box plots of (a) dentate and (b) interposed nuclei segmentation for ablation study. Statistical significance is marked as * and **, respectively, for $p<0.05$ and $p<0.001$. See Table S-V of the supplementary material for average values and deviation.

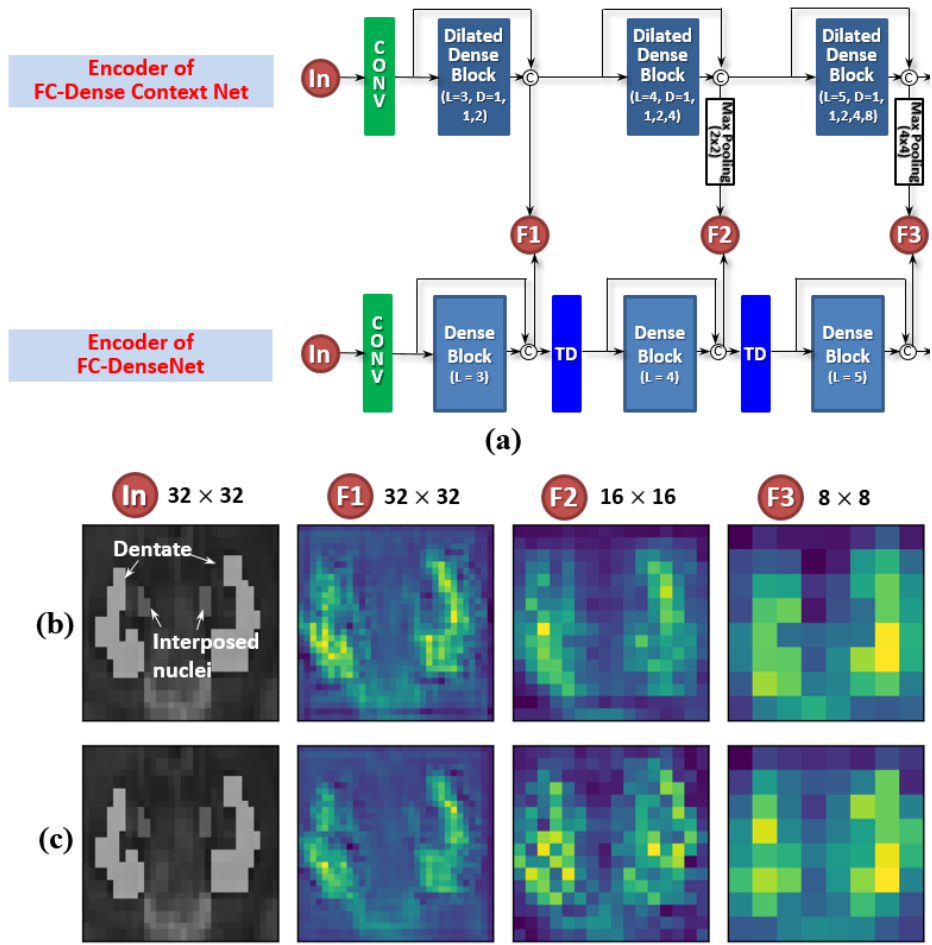

Fig. 12. Comparison of intermediate feature maps in encoders of FC-Dense ContextNet and FC-DenseNet. (a) Feature maps in each resolution (F1, F2, and F3) are extracted from skip-connection in the encoder of FC-Dense ContextNet (top) and FC-DenseNet (bottom) and convolution with 1×1 filter size is performed to visualize one channel feature map for simplicity. One slice image that contains dentate and interposed nuclei and has the same size as image patch (32×32) is used for the experiment. The input image and intermediate feature maps in the encoder of (b) FC-Dense ContextNet and (c) FC-DenseNet are compared. Feature maps F2 and F3 are scaled for visual comparison in each resolution.

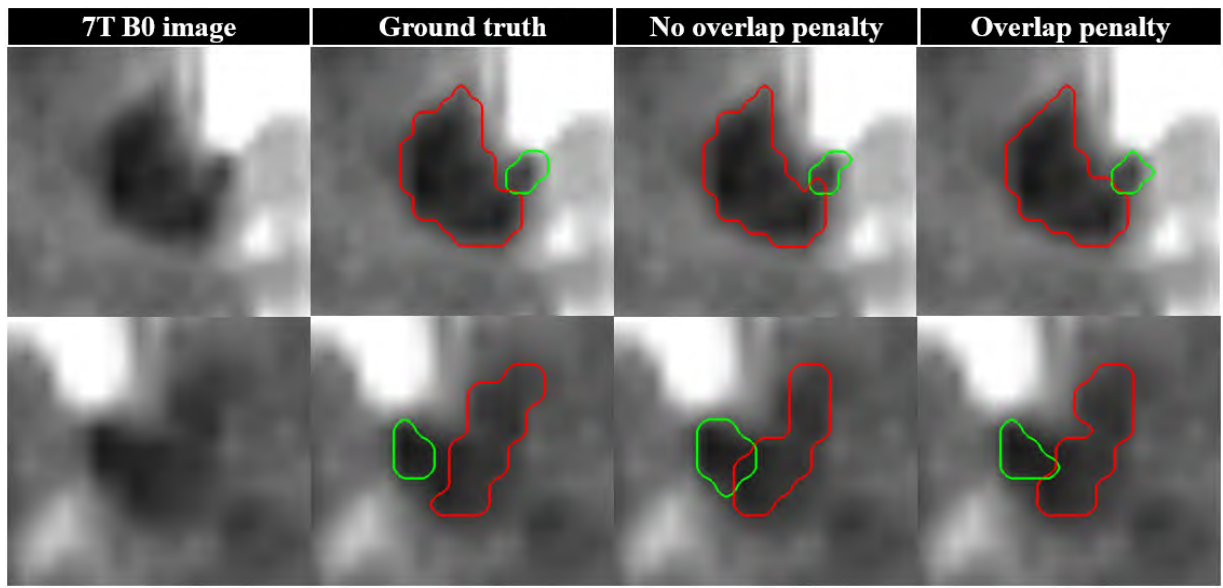

Fig. 13. A visual example on the axial plane that represents the effectiveness of overlap loss. Dentate (red) and interposed nuclei (green) of left (top) and right (bottom) are segmented using FC-Dense ContextNet (independent label estimation) without and with the overlap penalty.

TABLE I.
THE NUMBER OF TRAINABLE PARAMETERS FOR DEEP NEURAL NETWORKS

|  | U-Net | Attention U-Net | FC-DenseNet | DeepLab v3+ | CE-Net | DCN-Net |
|---|---|---|---|---|---|---|
| The number of trainable parameters | 5,605k | 5,862k | 742k | 1,280k | 2,742k | 890k |

TABLE II.

QUANTITATIVE COMPARISON BETWEEN JOINT AND INDEPENDENT LABEL PROBABILITY ESTIMATION IN FC-DENSE CONTEXTNET FOR DENTATE AND INTERPOSED NUCLEI SEGMENTATION.

| Target | Dentate | | | | Interposed | | | |
|---|---|---|---|---|---|---|---|---|
| Metric | CMD (mm) | MSD (mm$^2$) | DC | Volume (mm$^3$) | CMD (mm) | MSD (mm$^2$) | DC | Volume (mm$^3$) |
| Joint label estimation | 0.562±0.40 | 0.422±0.14 | 0.855±0.06 | 758±195 (668±159) | 1.141±0.88 | 0.488±0.25 | 0.669±0.14 | 55±18 (53±24) |
| Independent label estimation | 0.482±0.29 | **0.375±0.10** | **0.875±0.04** | 739±167 (668±159) | 1.115±0.89 | 0.489±0.31 | 0.673±0.14 | 59±21 (53±24) |

*Bold indicates p<0.001 for paired t-tests with joint label estimation. ( ) is ground truth volume.

TABLE III.

QUANTITATIVE COMPARISON OF FC-DENSE CONTEXTNET (INDEPENDENT LABEL PROBABILITY ESTIMATION) WITH ATTENTION LOSS AND OVERLAP LOSS IN DENTATE AND INTERPOSED NUCLEI SEGMENTATION.

| Target | Dentate | | | | Interposed | | | |
|---|---|---|---|---|---|---|---|---|
| Metric | CMD (mm) | MSD (mm$^2$) | DC | Volume (mm$^3$) | CMD (mm) | MSD (mm$^2$) | DC | Volume (mm$^3$) |
| FC-Dense ContextNet (independent label estimation) | 0.482±0.29 | 0.375±0.10 | 0.875±0.04 | 739±167 (668±159) | 1.115±0.89 | 0.489±0.31 | 0.673±0.14 | 59±21 (53±24) |
| + attention loss | 0.549±0.42 | 0.392±0.08 | **0.867±0.04** | 770±165 (668±159) | 1.040±0.75 | 0.472±0.24 | 0.680±0.12 | 59±23 (53±24) |
| + overlap loss | **0.555±0.37** | 0.377±0.10 | 0.874±0.04 | 735±172 (668±159) | 1.047±0.82 | 0.467±0.21 | 0.687±0.13 | 58±21 (53±24) |
| + attention loss and overlap loss (DCN-Net) | 0.514±0.35 | 0.380±0.11 | 0.873±0.05 | 736±165 (668±159) | 1.085±0.92 | 0.514±0.35 | 0.682±0.16 | 53±18 (53±24) |

*Bold indicates p<0.001 for paired t-tests with FC-Dense ContextNet. ( ) is ground truth volume.

TABLE IV.
QUANTITATIVE COMPARISON OF DCN-NET WITH SELF-TRAINING STRATEGIES IN DENTATE AND
INTERPOSED NUCLEI SEGMENTATION.

| Target | Dentate | | | | Interposed | | | |
|---|---|---|---|---|---|---|---|---|
| Metric | CMD (mm) | MSD (mm$^2$) | DC | Volume (mm$^3$) | CMD (mm) | MSD (mm$^2$) | DC | Volume (mm$^3$) |
| Manual labels | 0.514±0.35 | 0.380±0.11 | 0.873±0.05 | 736±165 (668±159) | 1.085±0.92 | 0.514±0.35 | 0.682±0.16 | 53±18 (53±24) |
| Self-training strategy #1 (pre-training) | **0.613±0.46** | 0.373±0.12 | 0.872±0.05 | 713±155 (668±159) | 1.029±0.76 | 0.486±0.22 | 0.676±0.15 | 52±16 (53±24) |
| Self-training strategy #1 (fine-tune) | 0.529±0.42 | 0.381±0.12 | 0.874±0.05 | 731±165 (668±159) | 1.003±0.78 | **0.424±0.24** | 0.686±0.14 | 67±24 (53±24) |
| Self-training strategy #2 (model distillation) | 0.552±0.44 | 0.390±0.13 | 0.868±0.05 | 778±185 (668±159) | 1.005±0.94 | 0.468±0.43 | 0.701±0.16 | 60±23 (53±24) |

*Bold indicates p<0.05 for paired t-tests with DCN-Net (trained on manual labels). ( ) is ground truth volume.

TABLE V.
QUANTITATIVE COMPARISON OF U-NET WITH SELF-TRAINING STRATEGIES IN DENTATE AND
INTERPOSED NUCLEI SEGMENTATION.

| Target | Dentate | | | | Interposed | | | |
|---|---|---|---|---|---|---|---|---|
| Metric | CMD (mm) | MSD (mm$^2$) | DC | Volume (mm$^3$) | CMD (mm) | MSD (mm$^2$) | DC | Volume (mm$^3$) |
| Manual labels | 0.616±0.53 | 0.431±0.13 | 0.853±0.06 | 724±184 (668±159) | 1.148±0.95 | 0.574±0.43 | 0.659±0.15 | 49±18 (53±24) |
| Self-training strategy #1 (pre-training) | **1.45±1.25** | **0.770±0.18** | **0.698±0.12** | 1216±267 (668±159) | 1.000±1.01 | 0.598±0.60 | 0.688±0.17 | 42±14 (53±24) |
| Self-training strategy #1 (fine-tune) | 0.668±0.56 | **0.462±0.17** | **0.837±0.08** | 793±213 (668±159) | **0.977±0.78** | 0.511±0.36 | **0.687±0.15** | 51±19 (53±24) |
| Self-training strategy #2 (model distillation) | 0.684±0.61 | 0.452±0.13 | 0.843±0.06 | 806±190 (668±159) | 1.02±0.93 | 0.52±0.38 | **0.693+0.17** | 48±20 (53±24) |

*Bold indicates p<0.05 for paired t-tests with U-Net (trained on manual labels). ( ) is ground truth volume.

# Supplementary Materials

We provide average values and deviation for each metric in addition to box plots displayed in Figures 6, 7, 8, and 11. Table S-I summarizes quantitative results of dentate and interposed nuclei segmentation obtained by using SUIT and DCN-Net. Quantitative results of dentate and interposed nuclei segmentation obtained by using state-of-the-art networks and DCN-Net with image patches of 5×5×5 and 9×9×9 step sizes, respectively, are summarized in Tables S-II and S-III. Also, Table S-IV presents quantitative results of dentate and interposed nuclei segmentation obtained by using image patches with different size within the proposed DCN-Net. Finally, quantitative results of dentate and interposed nuclei segmentation for ablation study are presented in Table S-V.

TABLE S-I.
QUANTITATIVE RESULTS OF DENTATE AND INTERPOSED NUCLEI SEGMENTATION OBTAINED BY USING SUIT AND DCN-NET IN COMPARISON TO GROUND TRUTH.

| Target | Dentate | | | | Interposed | | | |
|---|---|---|---|---|---|---|---|---|
| Metric | CMD (mm) | MSD (mm$^2$) | DC | Volume (mm$^3$) | CMD (mm) | MSD (mm$^2$) | DC | Volume (mm$^3$) |
| SUIT (DARTEL) | **2.437±1.65**[**] | **1.574±1.15**[**] | **0.626±0.20**[**] | **1153±463**[**] (668±159) | **1.912±1.30**[**] | **1.023±0.77**[**] | **0.448±0.23**[**] | **63±51**[**] (53±24) |
| DCN-Net | 0.514±0.35 | 0.38±0.11 | 0.873±0.05 | **736±166**[**] (668±159) | 1.085±0.92 | 0.514±0.35 | 0.682±0.16 | 53±18 (53±24) |

*Paired t-tests between SUIT and DCN-Net are performed for CMD, MSD, and DC. ANOVA and Tukey's post-hoc test are performed among volumes of SUIT segmentation, DCN-Net based segmentation, and ground truth. ( ) is the ground truth volume. Bold * and ** indicate p<0.05 and p<0.001, respectively.

TABLE S-II.

QUANTITATIVE RESULTS OF DENTATE AND INTERPOSED NUCLEI SEGMENTATION OBTAINED BY USING STATE-OF-THE-ART NETWORKS AND DCN-NET WITH IMAGE PATCHES OF 5×5×5 STEP SIZE IN COMPARISON TO GROUND TRUTH.

| Target | Dentate | | | | Interposed | | | |
|---|---|---|---|---|---|---|---|---|
| Metric | CMD (mm) | MSD (mm$^2$) | DC | Volume (mm$^3$) | CMD (mm) | MSD (mm$^2$) | DC | Volume (mm$^3$) |
| U-Net | **0.616±0.53**[*] | **0.431±0.13**[**] | **0.853±0.06**[**] | **724±184**[*] (668±159) | 1.148±0.95 | 0.574±0.43 | 0.659±0.15 | 49±18 (53±24) |
| Attention U-Net | 0.547±0.35 | **0.424±0.14**[**] | **0.858±0.06**[**] | **725±202**[**] (668±159) | 1.179±0.97 | 0.612±0.45 | 0.652±0.16 | 51±18 (53±24) |
| FC-DenseNet | **0.645±0.44**[*] | **0.446±0.12**[**] | **0.843±0.05**[**] | **819±205**[**] (668±159) | 1.149±0.84 | 0.535±0.39 | 0.658±0.16 | 56±22 (53±24) |
| Deeplab v3+ | **0.674±0.44**[**] | **0.455±0.12**[**] | **0.833±0.06**[**] | **842±197**[**] **(668±159)** | 1.118±0.84 | 0.534±0.36 | 0.67±0.14 | 57±22 (53±24) |
| CE-Net | **0.657±0.59**[*] | **0.476±0.30**[*] | **0.850±0.62**[**] | **736±184**[**] (668±159) | 1.244±0.99 | 0.577±0.38 | **0.649±0.15**[*] | 51±16 (53±24) |
| DCN-Net | 0.514±0.35 | 0.38±0.11 | 0.873±0.05 | **736±166**[**] (668±159) | 1.085±0.92 | 0.514±0.35 | 0.682±0.16 | 53±18 (53±24) |

*ANOVA and Tukey's post-hoc test among state-of-the-art networks and DCN-Net are performed for CMD, MSD, and DC. ANOVA and Tukey's post-hoc test are performed among volumes of state-of-the-art networks and DCN-Net based segmentation and ground truth. ( ) is the ground truth volume. Bold * and ** indicate p<0.05 and p<0.001, respectively, in comparison with DCN-Net.

TABLE S-III.

QUANTITATIVE RESULTS OF DENTATE AND INTERPOSED NUCLEI SEGMENTATION OBTAINED BY USING STATE-OF-THE-ART NETWORKS AND DCN-NET WITH IMAGE PATCHES OF 9×9×9 STEP SIZE IN COMPARISON TO GROUND TRUTH.

| Target | Dentate | | | | Interposed | | | |
|---|---|---|---|---|---|---|---|---|
| Metric | CMD (mm) | MSD (mm$^2$) | DC | Volume (mm$^3$) | CMD (mm) | MSD (mm$^2$) | DC | Volume (mm$^3$) |
| U-Net | 0.632±0.47 | **0.426±0.11**[*] | **0.856±0.05**[*] | 694±165 (668±159) | 1.27±1.00 | **0.618±0.45**[*] | **0.645±0.16**[*] | 48±17 (53±24) |
| Attention U-Net | 0.665±0.54 | **0.487±0.16**[**] | **0.833±0.06**[**] | **760±219**[**] (668±159) | 1.198±0.98 | **0.621±0.45**[*] | 0.661±0.15 | **46±17**[*] (53±24) |
| FC-DenseNet | **0.754±0.47**[*] | **0.462±0.11**[**] | **0.825±0.06**[**] | **841±200**[**] (668±159) | 1.375±1.60 | 0.760±1.28 | **0.635±0.18**[*] | **59±26**[*] (53±24) |
| Deeplab v3+ | **0.743±0.41**[*] | **0.469±0.11**[**] | **0.822±0.06**[**] | **845±192**[**] **(668±159)** | **1.270±0.90**[*] | 0.482±0.38 | 0.662±0.15 | **70±29**[**] (53±24) |
| CE-Net | **0.753±0.60**[*] | **0.535±0.11**[**] | **0.814±0.06**[**] | **866±216**[**] (668±159) | 1.271±0.93 | **0.659±0.44**[**] | **0.626±0.15**[*] | 50±17 (53±24) |
| DCN-Net | 0.593±0.46 | 0.398±0.09 | 0.868±0.04 | **731±167**[**] (668±159) | 1.093±0.84 | 0.451±0.24 | 0.676±0.15 | **62±22**[*] (53±24) |

*ANOVA and Tukey's post-hoc test among state-of-the-art networks and DCN-Net are performed for CMD, MSD, and DC. ANOVA and Tukey's post-hoc test are performed among volumes of state-of-the-art networks and DCN-Net based segmentation and ground truth. ( ) is the ground truth volume. Bold * and ** indicate p<0.05 and p<0.001, respectively, in comparison with DCN-Net.

TABLE S-IV.

QUANTITATIVE RESULTS OF DENTATE AND INTERPOSED NUCLEI SEGMENTATION OBTAINED BY USING DCN-NET WITH IMAGE PATCHES WITH DIFFERENT SIZES IN COMPARISON TO GROUND TRUTH.

| Target | Dentate | | | | Interposed | | | |
|---|---|---|---|---|---|---|---|---|
| Metric | CMD (mm) | MSD (mm$^2$) | DC | Volume (mm$^3$) | CMD (mm) | MSD (mm$^2$) | DC | Volume (mm$^3$) |
| 16×16×16 | 0.488±0.28 | 0.426±0.12 | 0.862±0.05 | **744±176**[**] (668±159) | 1.12±1.01 | 0.559±0.51 | 0.664±0.17 | **59±24**[*] (53±24) |
| 24×24×24 | 0.578±0.48 | 0.407±0.11 | 0.865±0.05 | **745±172**[**] (668±159) | 1.095±0.83 | 0.505±0.37 | 0.676±0.15 | **59±20**[*] (53±24) |
| 32×32×32 | **0.514±0.35**[*] | **0.38±0.11**[*] | **0.873±0.05**[*] | **736±166**[**] (668±159) | 1.085±0.92 | 0.514±0.35 | 0.682±0.16 | 53±18 (53±24) |
| 40×40×40 | 0.555±0.39 | 0.383±0.10 | 0.873±0.04 | **710±154**[**] (668±159) | 1.18±1.00 | 0.573±0.62 | 0.653±0.18 | **63±22**[**] (53±24) |

*Paired t-tests for each step are performed for CMD, MSD, and DC. ANOVA and Tukey's post-hoc test are performed among volumes of segmentation based on DCN-Net with different patch sizes and ground truth. ( ) is the ground truth volume. Bold * and ** indicate p<0.05 and p<0.001, respectively, in comparison with the previous patch size.

TABLE S-V.

QUANTITATIVE RESULTS OF DENTATE AND INTERPOSED NUCLEI SEGMENTATION FOR ABLATION STUDY IN COMPARISON TO GROUND TRUTH.

| Target | Dentate | | | | Interposed | | | |
|---|---|---|---|---|---|---|---|---|
| Metric | CMD (mm) | MSD (mm$^2$) | DC | Volume (mm$^3$) | CMD (mm) | MSD (mm$^2$) | DC | Volume (mm$^3$) |
| Dilated DenseNet | 0.692±0.49 | 0.498±0.14 | 0.815±0.07 | **897±214**[**] (668±159) | 1.219±0.91 | 0.534±0.30 | 0.634±0.17 | **64±22**[**] (53±24) |
| FC-DenseNet | 0.645±0.44 | **0.446±0.12**[*] | **0.843±0.05**[**] | **819±205**[**] (668±159) | 1.149±0.84 | 0.535±0.39 | 0.658±0.16 | 56±22 (53±24) |
| FC-Dense ContextNet (backbone) | 0.562±0.40 | 0.422±0.14 | **0.855±0.06**[*] | **758±195**[**] (668±159) | 1.141±0.88 | 0.488±0.25 | 0.669±0.14 | 55±18 (53±24) |
| DCN-Net | 0.514±0.35 | **0.380±0.11**[**] | **0.873±0.05**[**] | **736±166**[**] (668±159) | 1.085±0.92 | 0.514±0.35 | 0.682±0.16 | 53±18 (53±24) |
| DCN-Net (self-training #1) | 0.529±0.42 | 0.381±0.12 | 0.874±0.05 | **731±165**[**] (668±159) | 1.003±0.78 | **0.424±0.24**[*] | 0.686±0.14 | **67±24**[**] (53±24) |
| DCN-Net (self-training #2) | 0.552±0.44 | 0.390±0.13 | 0.868±0.05 | **777±185**[**] (668±159) | 1.005±0.94 | 0.468±0.44 | 0.701±0.16 | **60±23**[*] (53±24) |

*Paired t-tests for each step are performed for CMD, MSD, and DC. ANOVA and Tukey's post-hoc test are performed among volumes of segmentation based on networks with each component, DCN-Net based segmentation, and ground truth. ( ) is the ground truth volume. Bold * and ** indicate p<0.05 and p<0.001, respectively, in comparison with the network without each component.

Training loss and validation loss curve for the number of epochs in the backbone network (FC-Dense ContextNet) with the proposed multi-class hybrid asymmetric loss and DC loss, respectively, is presented in Fig. S-1. Training the network with DC loss was early stopped without convergence of the training and validation loss (at the stopping criteria of 20 epochs), resulting in segmentation failure, while training the network with the proposed multi-class hybrid asymmetric loss converged.

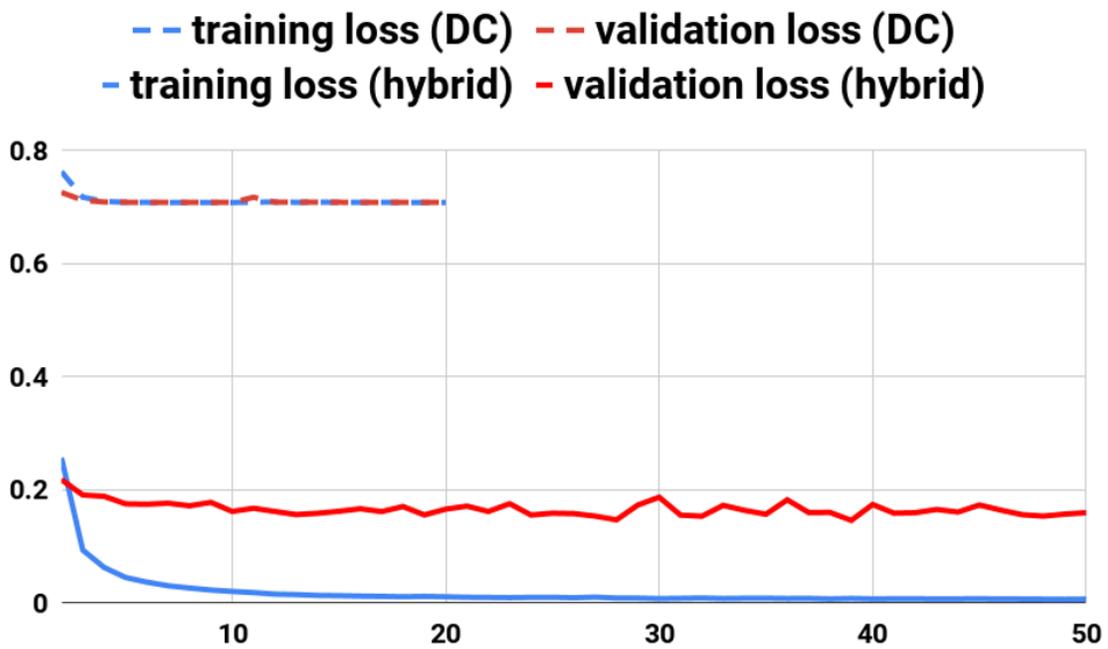

Fig. S-1. Training loss and validation loss curve using the proposed hybrid loss and DC loss in FC-Dense ContextNet.